\documentclass[final]{cvpr}
\usepackage{times}
\usepackage{epsfig}
\usepackage{graphicx}
\usepackage{amsmath}
\usepackage{amssymb}
\usepackage{xcolor}
\usepackage{multirow}
\usepackage{multicol}
\usepackage{subfigure, enumitem}
\usepackage{threeparttable, booktabs}

\usepackage[pagebackref=true,breaklinks=true,colorlinks,bookmarks=false]{hyperref}

\newcommand\blfootnote[1]{%
  \begingroup
  \renewcommand\thefootnote{}\footnote{#1}%
  \addtocounter{footnote}{-1}%
  \endgroup
}
\newcommand{\vect}[1]{\boldsymbol{#1}}

\DeclareMathOperator*{\argmin}{arg\,min}
\def\cvprPaperID{****} 

\begin{document}

\title{NTIRE 2021 Challenge on Quality Enhancement of Compressed Video: \\ Methods and Results}

\author{Ren Yang\and Radu Timofte\and 
Jing Liu \and Yi Xu \and Xinjian Zhang\and Minyi Zhao\and Shuigeng Zhou\and  
Kelvin C.K. Chan\and Shangchen Zhou\and Xiangyu Xu\and Chen Change Loy
\and Xin Li\and Fanglong Liu\and He Zheng\and Lielin Jiang\and Qi Zhang\and Dongliang He\and Fu Li\and Qingqing Dang
\and Yibin Huang\and Matteo Maggioni\and Zhongqian Fu\and Shuai Xiao\and Cheng li\and Thomas Tanay\and Fenglong Song
\and Wentao Chao \and Qiang Guo\and Yan Liu\and Jiang Li\and Xiaochao Qu\and
Dewang Hou\and Jiayu Yang\and Lyn Jiang\and Di You\and Zhenyu Zhang\and Chong Mou\and 
Iaroslav Koshelev\and Pavel Ostyakov\and Andrey Somov\and Jia Hao \and Xueyi Zou \and 
Shijie Zhao\and Xiaopeng Sun\and Yiting Liao\and Yuanzhi Zhang\and Qing Wang\and Gen Zhan\and Mengxi Guo\and Junlin Li \and
Ming Lu \and Zhan Ma \and 
Pablo Navarrete Michelini  \and 
Hai Wang \and Yiyun Chen\and Jingyu Guo\and Liliang Zhang\and Wenming Yang \and 
Sijung Kim \and Syehoon Oh \and 
Yucong Wang \and Minjie Cai \and 
Wei Hao \and Kangdi Shi\and Liangyan Li\and Jun Chen\and 
Wei Gao \and Wang Liu\and Xiaoyu Zhang\and  Linjie Zhou\and Sixin Lin\and Ru Wang 
\and
}

\maketitle
\thispagestyle{empty}
\pagestyle{empty}

\begin{abstract}
This paper reviews the first NTIRE challenge on quality enhancement of compressed video, with a focus on the proposed methods and results. In this challenge, the new Large-scale Diverse Video (LDV) dataset is employed. The challenge has three tracks. Tracks 1 and 2 aim at enhancing the videos compressed by HEVC at a fixed QP, while Track 3 is designed for enhancing the videos compressed by x265 at a fixed bit-rate. Besides, the quality enhancement of Tracks 1 and 3 targets at improving the fidelity (PSNR), and Track 2 targets at enhancing the perceptual quality. The three tracks totally attract 482 registrations. In the test phase, 12 teams, 8 teams and 11 teams submitted the final results of Tracks 1, 2 and 3, respectively. The proposed methods and solutions gauge the state-of-the-art of video quality enhancement. The homepage of the challenge: \url{https://github.com/RenYang-home/NTIRE21_VEnh}
\end{abstract}

\vspace{-1em}
\section{Introduction}
\vspace{-1em}
\blfootnote{Ren Yang ({\tt ren.yang@vision.ee.ethz.ch}, ETH Z\"urich) and Radu Timofte ({\tt radu.timofte@vision.ee.ethz.ch}, ETH Z\"urich) are the organizers of the NTIRE 2021 challenge, and other authors participated in the challenge.\\
The Appendix lists the authors’ teams and affiliations.\\
NTIRE 2021 website: \url{https://data.vision.ee.ethz.ch/cvl/ntire21/}}

The recent years have witnessed the increasing popularity of video streaming over the Internet~\cite{Cisco} and meanwhile the demands on high-quality and high-resolution videos are also increasing. To transmit larger number of high-resolution videos through the bandwidth-limited Internet, video compression~\cite{wiegand2003overview, sullivan2012overview} has to be applied to significantly reduce the bit-rate. However, compression artifacts unavoidably occur in compressed videos, and may lead to severe quality degradation. Therefore, it is essential to study on improving the quality of compressed video. The NTIRE 2021 Challenge aims at establishing the benchmarks on enhancing compressed video towards both fidelity and perceptual quality. 

In the past a few years, there has been plenty of works in this direction~\cite{yang2017decoder, yang2018enhancing, wang2017novel, lu2018deep, yang2018multi, xu2021multi, yang2019quality, guan2019mfqe, Xu_2019_ICCV, deng2020spatio, yang2020learning, huo2021recurrent, wang2020multi}, among which~\cite{yang2017decoder, yang2018enhancing, wang2017novel} are single-frame quality enhancement methods, while~\cite{lu2018deep, yang2018multi, yang2019quality, guan2019mfqe, Xu_2019_ICCV, deng2020spatio, yang2020learning, huo2021recurrent, wang2020multi} propose enhancing quality by taking advantage of temporal correlation. Besides, \cite{wang2020multi} aims at improving the perceptual quality of compressed video. Other works~\cite{yang2017decoder, yang2018enhancing, wang2017novel, lu2018deep, yang2018multi, yang2019quality, guan2019mfqe, Xu_2019_ICCV, deng2020spatio, yang2020learning, huo2021recurrent} focus on advancing the performance on Peak Signal-to-Noise Ratio (PSNR) to achieve higher fidelity to the uncompressed video. These works show the promising future of this research field. However, as discussed in \cite{yang2021dataset}, the scales of the training sets in previous methods are incremental and different methods are also tested on different test sets. 

The NTIRE 2021 challenge on quality enhancement of compressed video is a step forward for establishing a benchmark of video quality enhancement algorithms. It uses the newly proposed Large-scale Diverse Video (LDV)~\cite{yang2021dataset} dataset, which contains 240 videos with the diversities of content, motion, frame-rate, \etc. The LDV dataset is introduced in \cite{yang2021dataset} along with the analyses of the challenge results. In the following, we first describe the NTIRE 2021 challenge, and then introduce the proposed methods and their results.

\section{NTIRE 2021 Challenge}

The objectives of the NTIRE 2021 challenge on enhancing compressed video are: (i) to advance the state-of-the-art in video quality enhancement; (ii) to compare different solutions; (iii) to promote the newly proposed LDV dataset; and (iv) to study quality enhancement on more challenging video compression settings.

This challenge is one of the NTIRE 2021 associated challenges: nonhomogeneous dehazing~\cite{ancuti2021ntire}, defocus deblurring using dual-pixel~\cite{abuolaim2021ntire}, depth guided image relighting~\cite{elhelou2021ntire}, image deblurring~\cite{nah2021ntire}, multi-modal aerial view imagery classification~\cite{liu2021ntire}, learning the super-resolution space~\cite{lugmayr2021ntire}, quality enhancement of compressed video (this report), video super-resolution~\cite{son2021ntire}, perceptual image quality assessment~\cite{gu2021ntire}, burst super-resolution~\cite{bhat2021ntire}, and high dynamic range imaging~\cite{perez2021ntire}.

\subsection{LDV dataset}

As introduced in \cite{yang2021dataset}, our LDV dataset contains 240 videos with 10 categories of scenes, \ie, \textit{animal}, \textit{city}, \textit{close-up}, \textit{fashion}, \textit{human}, \textit{indoor}, \textit{park}, \textit{scenery}, \textit{sports} and \textit{vehicle}. Besides, among the 240 videos in LDV, there are 48 fast-motion videos, 68 high frame-rate ($\geq50$) videos and 172 low frame-rate ($\leq30$) videos. Additionally, the camera is slightly shaky (\eg, captured by handheld camera) in 75 videos of LDV, and 20 videos in LDV are with the dark environments, \eg, at night or in the rooms with insufficient light. In the challenge of NTIRE 2021, we divide the LDV dataset into training, validation and test sets with 200, 20 and 20 videos, respectively. The test set is further split into two sets with 10 videos in each for the tracks of fixed QP (Tracks 1 and 2) and fixed bit-rate (Track 3), respectively. The 20 validation videos consist of the videos from the 10 categories of scenes with two videos in each category. Each test set has one video from each category. Besides, 9 out of the 20 validation videos and 4 among the 10 videos in each test set are with high frame-rates. There are five fast-motion videos in the validation set. In the test sets for fixed QP and fixed bit-rate tracks, there are three and two fast-motion videos, respectively.  

\begin{table*}[!t]
\footnotesize
  \centering
  \caption{The results of Track 1 (fixed QP, fidelity)}
    \begin{tabular}{ccccccccccccc}
    \cmidrule[\heavyrulewidth]{1-13}
    \multirow{2}[3]{*}{Team} & \multicolumn{11}{c}{PSNR (dB)}                                                        &        \multicolumn{1}{c}{MS-SSIM} \\
\cmidrule(r){2-12}\cmidrule(l){13-13}          & \#1 & \#2 & \#3 & \#4 & \#5 & \#6 & \#7 & \#8 & \#9 & \#10 & Average &        \multicolumn{1}{c}{Average} \\
\cmidrule{1-13}    BILIBILI AI \& FDU & \textbf{33.69} & 31.80 & \textbf{38.31} & 34.44 & 28.00 & 32.13 & 29.68 & \textbf{29.91} & 35.61 & 31.62 & \textbf{32.52} &        \multicolumn{1}{c}{\textbf{0.9562}} \\
    NTU-SLab & 31.30 & \textbf{32.46} & 36.96 & \textbf{35.29} & \textbf{28.30} & \textbf{33.00} & \textbf{30.42} & 29.20 & \textbf{35.70} & \textbf{32.24} & 32.49 &        \multicolumn{1}{c}{0.9552} \\
    VUE   & 31.10 & 32.00 & 36.36 & 34.86 & 28.08 & 32.26 & 30.06 & 28.54 & 35.31 & 31.82 & 32.04 & \multicolumn{1}{c}{0.9493} \\
    NOAHTCV & 30.97 & 31.76 & 36.25 & 34.52 & 28.01 & 32.11 & 29.75 & 28.56 & 35.38 & 31.67 & 31.90 &        \multicolumn{1}{c}{0.9480} \\
    Gogoing & 30.91 & 31.68 & 36.16 & 34.53 & 27.99 & 32.16 & 29.77 & 28.45 & 35.31 & 31.66 & 31.86 &        \multicolumn{1}{c}{0.9472} \\
    NJU-Vision & 30.84 & 31.55 & 36.08 & 34.47 & 27.92 & 32.01 & 29.72 & 28.42 & 35.21 & 31.58 & 31.78 &        \multicolumn{1}{c}{0.9470} \\
    MT.MaxClear & 31.15 & 31.21 & 37.06 & 33.83 & 27.68 & 31.68 & 29.52 & 28.43 & 34.87 & 32.03 & 31.75 &      \multicolumn{1}{c}{0.9473}  \\
    VIP\&DJI & 30.75 & 31.36 & 36.07 & 34.35 & 27.79 & 31.89 & 29.48 & 28.35 & 35.05 & 31.47 & 31.65 &       \multicolumn{1}{c}{0.9452} \\
    Shannon & 30.81 & 31.41 & 35.83 & 34.17 & 27.81 & 31.71 & 29.53 & 28.43 & 35.05 & 31.49 & 31.62 &        \multicolumn{1}{c}{0.9457} \\
    HNU\_CVers & 30.74 & 31.35 & 35.90 & 34.21 & 27.79 & 31.76 & 29.49 & 28.24 & 34.99 & 31.47 & 31.59 &        \multicolumn{1}{c}{0.9443} \\
    BOE-IOT-AIBD & 30.69 & 30.95 & 35.65 & 33.83 & 27.51 & 31.38 & 29.29 & 28.21 & 34.94 & 31.29 & 31.37 &        \multicolumn{1}{c}{0.9431} \\
    Ivp-tencent & 30.53 & 30.63 & 35.16 & 33.73 & 27.26 & 31.00 & 29.22 & 28.14 & 34.51 & 31.14 & 31.13 &        \multicolumn{1}{c}{0.9405} \\
    \midrule
    \midrule
    MFQE~\cite{yang2018multi} & 30.56 & 30.67 & 34.99 & 33.59 & 27.38 & 31.02 & 29.21 & 28.03 & 34.63 & 31.17 & 31.12 &  0.9392  \\
    QECNN~\cite{yang2018enhancing} & 30.46 & 30.47 & 34.80 & 33.48 & 27.17 & 30.78 & 29.15 & 28.03 & 34.39 & 31.05 & 30.98 & 0.9381 \\
    DnCNN~\cite{zhang2017beyond} & 30.41 & 30.40 & 34.71 & 33.35 & 27.12 & 30.67 & 29.13 & 28.00 & 34.37 & 31.02 & 30.92 & 0.9373 \\
    ARCNN~\cite{dong2015compression} & 30.29 & 30.18 & 34.35 & 33.12 & 26.91 & 30.42 & 29.05 & 27.97 & 34.23 & 30.87 & 30.74 & 0.9345 \\

    \midrule
    \midrule
    Unprocessed video & 30.04 & 29.95 & 34.16 & 32.89 & 26.79 & 30.07 & 28.90 & 27.84 & 34.09 & 30.71 & 30.54 &   \multicolumn{1}{c}{0.9305}      \\
    \cmidrule[\heavyrulewidth]{1-13}
    \end{tabular}%
  \label{tab:track1}%
\end{table*}%

\subsection{Fidelity tracks}

The first part of this challenge aims at improving the quality of compressed video towards fidelity. We evaluate the fidelity via PSNR. Additionally, we also calculate the Multi-Scale Structural SIMilarity index (MS-SSIM)~\cite{wang2003multiscale} for the proposed methods.

\textbf{Track 1: Fixed QP. } In Track 1, the videos are compressed following the typical settings of the existing literature~\cite{yang2017decoder, yang2018enhancing, wang2017novel, lu2018deep, yang2018multi, yang2019quality, guan2019mfqe, Xu_2019_ICCV, deng2020spatio, huo2021recurrent, wang2020multi}, \ie, using the official HEVC test model (HM) at fixed QPs. In this challenge, we compress videos by the default configuration of the Low-Delay P (LDP) mode (\textit{encoder\_lowdelay\_P\_main.cfg}) of HM 16.20\footnote{\url{https://hevc.hhi.fraunhofer.de/svn/svn_HEVCSoftware/tags/HM-16.20}} at QP = 37. In this setting, due to the regularly changed QPs at the frame level, the compression quality normally fluctuates regularly among frames. Besides, it does not enable the rate control strategy, and therefore the frame-rate does not have impact on compression. This may make this track to be an easy task.

\textbf{Track 3: Fixed bit-rate. } Track 3 targets at a more practical scenario. In video streaming, rate control has been utilizing as a popular strategy to constraint the bit-rate into the limited bandwidth. In this track, we compress videos by the x265 library of FFmpeg\footnote{\url{https://johnvansickle.com/ffmpeg/releases/ffmpeg-release-amd64-static.tar.xz}} with rate control enabled and set the target bit-rate as 200 kbps, by the following commands:

\

\noindent\texttt{ffmpeg -pix\_fmt yuv420p -s WxH -r FR -i name.yuv -c:v libx265 -b:v 200k -x265-params pass=1:log-level=error -f null /dev/null}

\

\noindent\texttt{ffmpeg -pix\_fmt yuv420p -s WxH -r FR -i name.yuv -c:v libx265 -b:v 200k -x265-params pass=2:log-level=error name.mkv}

\

\noindent Note that we utilize the two-pass scheme to ensure the accuracy of rate control. Due to the fixed bit-rate, the videos of different frame-rates, various motion speeds and diverse contents have to be compressed to a specific bit-rate per second. This makes the compression quality of different videos dramatically different, and therefore may make it a more challenging track than Track 1.

\subsection{Perceptual track}

We also organize a track aiming at enhancing the compressed videos towards perceptual quality. In this track, the performance is evaluated via the Mean Opinion Score (MOS)~\cite{mos}. We also report the performance on other perceptual metrics as references, such as the Learned Perceptual Image Patch Similarity (LPIPS)~\cite{zhang2018unreasonable}, Fr\'echet Inception Distance (FID)~\cite{heusel2017gans}, Kernel Inception Distance (KID)~\cite{binkowski2018demystifying} and Video Multimethod Assessment Fusion (VMAF)~\cite{VMAF}.

\textbf{Track 2: perceptual quality enhancement. } In Track~2, we compress videos with the same settings as Track~1. The task of this track is to generate visually pleasing enhanced videos, and the scores are ranked according to MOS values~\cite{mos} from 15 subjects. The scores range from $s=0$ (poorest quality) to $s=100$ (best quality). The groundtruth videos are given to the subjects as the standard of $s=100$, but the subjects are asked to rate videos in accordance with the visual quality, instead of the similarity to the groundtruth. We linearly normalize the scores ($s$) of each subject to 
\begin{equation}
    s' = 100 \cdot \frac{s-s_{min}}{s_{max} - s_{min}},
\end{equation}
in which $s_{max}$ and $s_{min}$ denote the highest and the lowest score of each subject, respectively. In the experiment, we insert five repeated videos to check the concentration of each subject to ensure the consistency of rating. Eventually, we omit the scores from the four least concentrated subjects whose average errors on repeated videos are larger than 20. Hence, the final MOS values are averaged among 11 subjects. Besides, we also calculate the LPIPS, FID, KID and VMAF values to evaluate the proposed methods.

\begin{table*}[!t]
\footnotesize
  \centering
  \caption{The results of Track 2 (fixed QP, perceptual)}
    \begin{tabular}{cccccccccccccccc}
    \toprule
    \multirow{2}[4]{*}{Team} & \multicolumn{11}{c}{MOS $\uparrow$}                                                              & LPIPS $\downarrow$ & FID $\downarrow$   & KID $\downarrow$   & VMAF $\uparrow$ \\
\cmidrule(r){2-12}\cmidrule(lr){13-13}\cmidrule(lr){14-14}\cmidrule(lr){15-15}\cmidrule(l){16-16}         & \#1   & \#2   & \#3   & \#4   & \#5   & \#6   & \#7   & \#8   & \#9   & \#10  & Average  & Average & Average & Average & Average \\
    \midrule

    BILIBILI AI \& FDU & \textbf{90} & \textbf{74} & 66    & 70    & \textbf{86} & 60    & 56    & \textbf{95} & 66    & 59    & \textbf{72} & \textbf{0.0429} & \textbf{32.17} & \textbf{0.0137} & 75.69 \\
    NTU-SLab & 63    & \textbf{74} & 65    & \textbf{74} & 74    & 65    & 61    & 80    & 64    & \textbf{81} & 70    & 0.0483 & 34.64 & 0.0179 & 71.55 \\
    NOAHTCV & 73    & \textbf{74} & 71    & 64    & 71    & \textbf{86} & \textbf{63} & 55    & 58    & 56    & 67    & 0.0561 & 46.39 & 0.0288 & 68.92 \\
    Shannon & 67    & 67    & 78    & 73    & 68    & 66    & 57    & 64    & 60    & 58    & 66    & 0.0561 & 50.61 & 0.0332 & 69.06 \\
    VUE   & 60    & 69    & 75    & 64    & 78    & 55    & 62    & 32    & 52    & 61    & 61    & 0.1018 & 72.27 & 0.0561 & \textbf{78.64} \\
    BOE-IOT-AIBD & 51    & 42    & 67    & 69    & 68    & 45    & 61    & 50    & 42    & 52    & 55    & 0.0674 & 62.05 & 0.0447 & 68.78 \\
    (\textit{anonymous}) & 46    & 41    & 69    & 59    & 63    & 31    & \textbf{63} & 21    & \textbf{70} & 47    & 51    & 0.0865 & 83.77 & 0.0699 & 69.70 \\
    MT.MaxClear & 33    & 50    & \textbf{81} & 65    & 52    & 40    & 56    & 14    & 41    & 31    & 46    & 0.1314 & 92.42 & 0.0818 & 77.30 \\
    \midrule
    \midrule
    Unprocessed video &   35    & 27    & 31    & 57    & 38    & 30    & 34    & 41    & 32    & 36    & 36  & 0.0752 & 48.94 & 0.0303 & 65.72
 \\
    \bottomrule
    \end{tabular}%
  \label{tab:track2}%
\end{table*}%

\begin{table*}[!t]
\footnotesize
  \centering
  \caption{The results of Track 3 (fixed bit-rate, fidelity)}
    \begin{tabular}{cccccccccccc@{\hskip 5ex}c}
    \cmidrule[\heavyrulewidth]{1-13}
    \multirow{2}[3]{*}{Team} & \multicolumn{11}{c}{PSNR (dB)}                                                        &        \multicolumn{1}{c}{MS-SSIM} \\
\cmidrule(r{.5em}){2-12}\cmidrule(l{.5em}){13-13}          & \#11 & \#12 & \#13 & \#14 & \#15 & \#16 & \#17 & \#18 & \#19 & \#20 & Average &        \multicolumn{1}{c}{Average} \\
\cmidrule{1-13}    
    
    NTU-SLab & \textbf{30.59} & \textbf{28.14} & 35.37 & \textbf{34.61} & \textbf{32.23} & \textbf{34.66} & 28.17 & 20.38 & 27.39 & \textbf{32.13} & \textbf{30.37} & \multicolumn{1}{c}{\textbf{0.9484}} \\
    BILIBILI AI \& FDU & 29.85 & 27.01 & 34.17 & 34.25 & 31.62 & 34.34 & \textbf{28.51} & \textbf{21.13} & \textbf{28.01} & 30.65 & 29.95 & \multicolumn{1}{c}{0.9468} \\
    MT.MaxClear & 29.47 & 27.89 & \textbf{35.63} & 34.16 & 30.93 & 34.29 & 26.25 & 20.47 & 27.38 & 30.38 & 29.69 & \multicolumn{1}{c}{0.9423} \\
    Block2Rock Noah-Hisilicon & 30.20 & 27.31 & 34.50 & 33.55 & 31.94 & 34.14 & 26.62 & 20.43 & 26.74 & 30.96 & 29.64 & \multicolumn{1}{c}{0.9405} \\
    VUE   & 29.93 & 27.31 & 34.58 & 33.64 & 31.79 & 33.86 & 26.54 & 20.44 & 26.54 & 30.97 & 29.56 & \multicolumn{1}{c}{0.9403} \\
    Gogoing & 29.77 & 27.23 & 34.36 & 33.47 & 31.61 & 33.71 & 26.68 & 20.40 & 26.38 & 30.77 & 29.44 & \multicolumn{1}{c}{0.9393} \\
    NOAHTCV & 29.80 & 27.13 & 34.15 & 33.38 & 31.60 & 33.66 & 26.38 & 20.36 & 26.37 & 30.64 & 29.35 & \multicolumn{1}{c}{0.9379} \\
    BLUEDOT & 29.74 & 27.09 & 34.08 & 33.29 & 31.53 & 33.33 & 26.50 & 20.36 & 26.35 & 30.57 & 29.28 & \multicolumn{1}{c}{0.9384} \\
    VIP\&DJI & 29.64 & 27.09 & 34.12 & 33.44 & 31.46 & 33.50 & 26.50 & 20.34 & 26.19 & 30.56 & 29.28 & \multicolumn{1}{c}{0.9380} \\
    McEhance & 29.57 & 26.81 & 33.92 & 33.10 & 31.36 & 33.40 & 25.94 & 20.21 & 26.07 & 30.27 & 29.07 & \multicolumn{1}{c}{0.9353} \\
    BOE-IOT-AIBD & 29.43 & 26.68 & 33.72 & 33.02 & 31.04 & 32.98 & 26.25 & 20.26 & 25.81 & 30.09 & 28.93 & \multicolumn{1}{c}{0.9350} \\
    \midrule
    \midrule
    Unprocessed video & 29.17 & 26.02 & 32.52 & 32.22 & 30.69 & 32.54 & 25.48 & 20.03 & 25.28 & 29.41 & 28.34 & \multicolumn{1}{c}{0.9243} \\
    \cmidrule[\heavyrulewidth]{1-13}
    \end{tabular}%
  \label{tab:track3}%
\end{table*}%

\section{Challenge results}

\subsection{Track 1: Fixed QP, Fidelity}

The numerical results of Track 1 are shown in Table~\ref{tab:track1}. On the top part, we show the results of the 12 methods proposed in this challenge. The unprocessed video indicates the compressed videos without enhancement. Additionally, we also train the models of the existing methods on the training set of the newly proposed LDV dataset, and report the results in Table~\ref{tab:track1}. It can be seen from Table~\ref{tab:track1}, the proposed methods in the challenge outperform the existing methods, and therefore advance the state-of-the-art of video quality enhancement.

The PSNR improvement of the 12 proposed methods ranges from 0.59 dB to 1.98 dB, and the improvement of MS-SSIM ranges between 0.0100 and 0.0257. The BILIBILI AI \& FDU Team achieves the best average PSNR and MS-SSIM performance in this track. They improve the average PSNR and MS-SSIM by 1.98 dB and 0.0257, respectively. The NTU-SLab and VUE Teams rank second and third, respectively. The average PSNR performance of NTU-SLab is slightly lower (0.03 dB) than the BILIBILI AI \& FDU Team, and the PSNR of VUE is 0.48 dB lower than the best method.
We also report the detailed results on the 10 test videos (\#1 to \#10) in Table~\ref{tab:track1}. The results indicate that the second-ranked team NTU-SLab outperforms BILIBILI AI \& FDU on 7 videos. It shows the better generalization capability of NTU-SLab than BILIBILI AI \& FDU. Considering the average PSNR and generalization, the BILIBILI AI \& FDU and NTU-SLab Teams are both the winners of this track.

\subsection{Track 2: Fixed QP, Perceptual}

Table~\ref{tab:track2} shows the results of Track 2. In Track 2, the BILIBILI AI \& FDU Team achieves the best MOS performance on 4 of the 10 test videos and has the best average MOS performance. The results of the NTU-SLab team are the best on 3 videos, and their average MOS performance ranks second. The NOAHTCV Team is the third in the ranking of average MOS. We also report the results of LIPIS, FID, KID and VMAF, which are the popular metrics for evaluating perceptual quality of image and video. It can be seen from Table~\ref{tab:track2} that BILIBILI AI \& FDU, NTU-SLab and NOAHTCV still rank at the first, second and third places on LPIPS, FID and KID. It indicates that these perceptual metrics are effective on measuring the subjective quality. However, the rank on VMAF is obviously different from MOS. Besides, some teams perform worse than the unprocessed videos on LPIPS, FID and KID, while their MOS values are all higher than the unprocessed videos. This may show that the perceptual metrics are not always reliable, and the metrics LPIPS, FID and KID, which are designed for image, may be not very suitable for evaluating the visual quality of video.

\subsection{Track 3: Fixed bit-rate, Fidelity}

Table~\ref{tab:track3} shows the results of Track 3. In this track, we use the different videos as the test set, denoted as \#11 to \#20. The top three teams in this track are NTU-SLab, BILIBILI AI \& FDU and MT.MaxClear. The NTU-SLab Team achieves the best results on 6 videos and also ranks first on average PSNR and MS-SSIM. They improve the average PSNR by 2.03 dB. BILIBILI AI \& FDU and MT.MaxClear enhance PSNR by 1.61 dB and 1.35 dB, respectively.

\begin{table*}[htbp]
\scriptsize
  \centering
  \caption{The reported time complexity, platforms, test strategies and training data of the challenge methods.}
    \begin{tabular}{cccccccc}
    \toprule
    \multirow{2}[4]{*}{Team} & \multicolumn{3}{c}{Running time (s) per frame} & \multirow{2}[4]{*}{Platform} & \multirow{2}[4]{*}{GPU} & \multirow{2}[4]{*}{Ensemble / Fusion} & \multirow{2}[4]{*}{Extra training data}\\
\cmidrule{2-4}          & Track 1 & Track 2 & Track 3 &       &       &  \\
    \midrule
    BILIBILI AI \& FDU & 9.00     & 9.45  & 9.00     & PyTorch & Tesla V100/RTX 3090 & Flip/Rotation  x8 & Bilibili~\cite{bili}, YouTube~\cite{youtube} \\
    NTU-SLab & 3.44  & 3.44  & 3.44  & PyTorch & Tesla V100 & Flip/Rotation  x8 & Pre-trained on REDS~\cite{nah2019ntire}\\
VUE   & 34    & 36    & 50    & PyTorch & Tesla V100 & Flip/Rotation  x8 & Vimeo90K~\cite{xue2019video} \\
    NOAHTCV & 12.8  & 12.8  & 12.8  & TensorFlow & Tesla V100 & Flip/Rotation  x8 & DIV8K~\cite{gu2019div8k} (Track 2)\\
    MT.MaxClear & 2.4   & 2.4   &  2.4    & PyTorch & Tesla V100 & Flip/Rotation/Multi-model x12 & Private dataset\\
    Shannon & 12.0   & 1.5   & -     & PyTorch & Tesla T4 & Flip/Rotation  x8 (Track 1) & -\\
    Block2Rock Noah-Hisilicon & -     & -     & 300    & PyTorch & Tesla V100 & Flip/Rotation  x8 & YouTube~\cite{youtube}\\
    Gogoing & 8.5   & -     & 6.8   & PyTorch & Tesla V100 & Flip/Rotation x4 & REDS~\cite{nah2019ntire}\\
    NJU-Vision & 4.0   & -     & -     & PyTorch & Titan RTX & Flip/Rotation x8 & SJ4K~\cite{song2013sjtu} \\
    BOE-IOT-AIBD & 1.16  & 1.16  & 1.16  & PyTorch & GTX 1080 & Overlapping patches & - \\
    (\textit{anonymous}) & -     & 4.52  & -     & PyTorch & Tesla V100 & - & Partly finetuned from \cite{wang2019edvr} \\
    VIP\&DJI & 18.4   & -     & 12.8  & PyTorch & GTX 1080/2080 Ti & Flip/Rotation  x8 & SkyPixel~\cite{SKYPIXEL}.\\
    BLUEDOT & -     & -     & 2.85  & PyTorch & RTX 3090 & - & Dataset of MFQE 2.0~\cite{guan2019mfqe}\\
    HNU\_CVers & 13.72 & -     & -     & PyTorch & RTX 3090 & Overlapping patches & - \\
    McEhance & -     & -     & 0.16  & PyTorch & GTX 1080 Ti & - & -\\
    Ivp-tencent & 0.0078  & -     & -     & PyTorch & GTX 2080 Ti & - & -\\
     \midrule
     \midrule
    MFQE~\cite{yang2018multi} & 0.38 & - & - & TensorFlow & TITAN Xp & - & -\\
    QECNN~\cite{yang2018enhancing}  & 0.20 & - & - & TensorFlow & TITAN Xp & - & -\\
    DnCNN~\cite{zhang2017beyond}  & 0.08 & - & - & TensorFlow & TITAN Xp & - & -\\
    ARCNN~\cite{dong2015compression}  & 0.02 & - & - & TensorFlow & TITAN Xp & - & -\\
    \bottomrule
    \end{tabular}%
  \label{tab:time}%
\end{table*}%

\subsection{Efficiency}

Table~\ref{tab:time} reports the running time of the proposed methods. The NTU-SLab and BILIBILI AI \& FDU Teams achieve the best and the second best quality performance for all three tracks, while NTU-SLab is several times faster than BILIBILI AI \& FDU. Therefore, the NTU-SLab Team makes a good trade-off between quality and time efficiency. In Track 3, the MT.MaxClear Team has the fastest speed among the top 3 methods. Moreover, Ivp-tencent is most time-efficient in all proposed methods. It is able to enhance more than 120 frames per second, so it may be practical for the scenario of high frame-rates. Note that the running time is tested on the machine of each team and the numbers are reported by the teams' authors. Therefore, the values in Table~\ref{tab:time} are only for reference, since it is hard to guarantee the fairness for comparing time efficiency. 

\subsection{Training and test}

It can be seen from Table~\ref{tab:time} that the top teams utilized extra training data in additional to the 200 training videos of LDV~\cite{yang2021dataset} provided in the challenge. It may indicate that the scale of training database has obvious effect on the test performance.
Besides, the ensemble strategy~\cite{timofte2016seven} has been widely used in the top methods, and the participants observe the quality improvement of their methods when using ensemble, showing the effectiveness of the ensemble strategy for video enhancement.

\begin{figure}[!t]
	\centering
	\subfigure[Network architecture.]{\includegraphics[width=\linewidth]{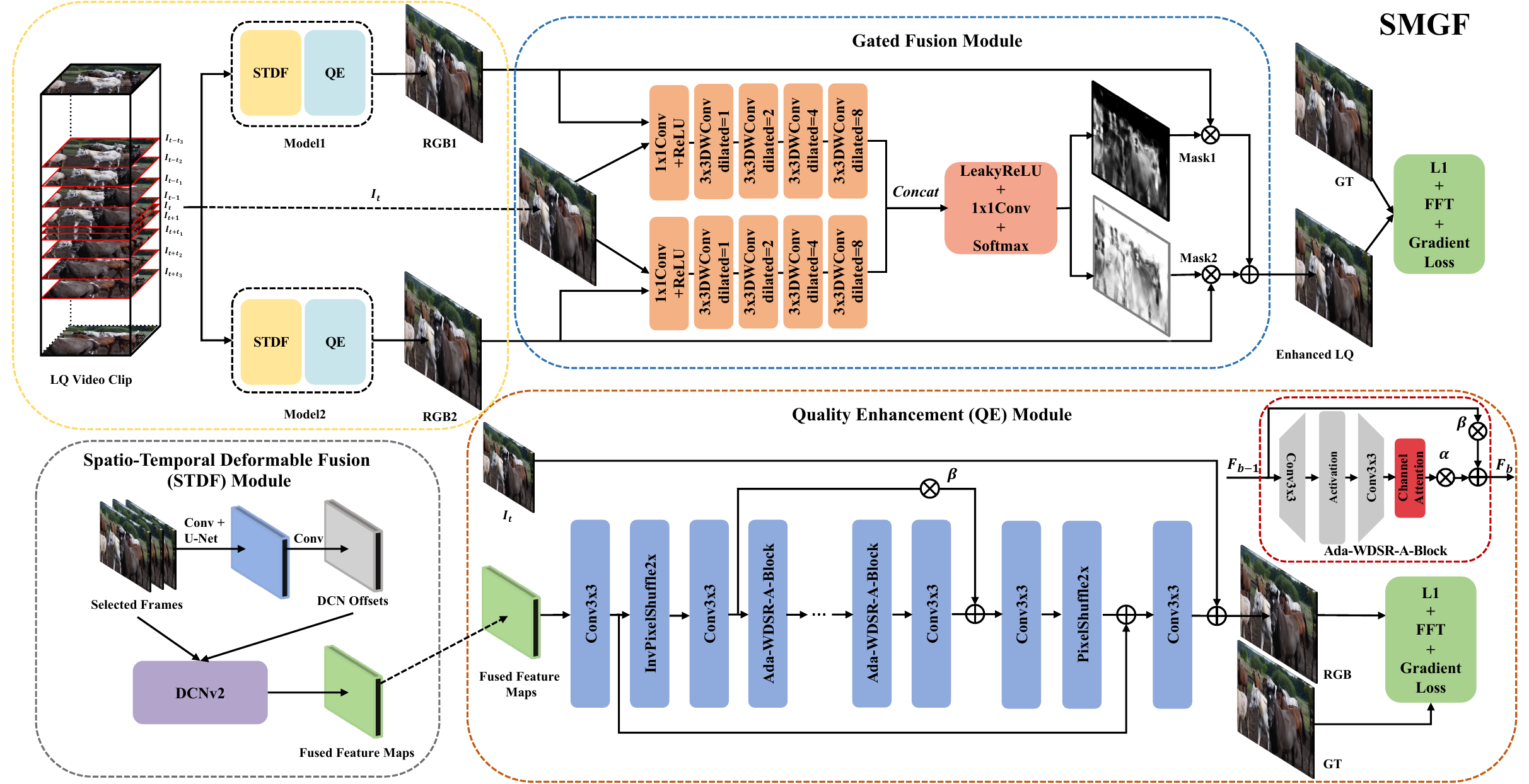}}\\
	\subfigure[The proposed pSMGF.]{\includegraphics[width=\linewidth]{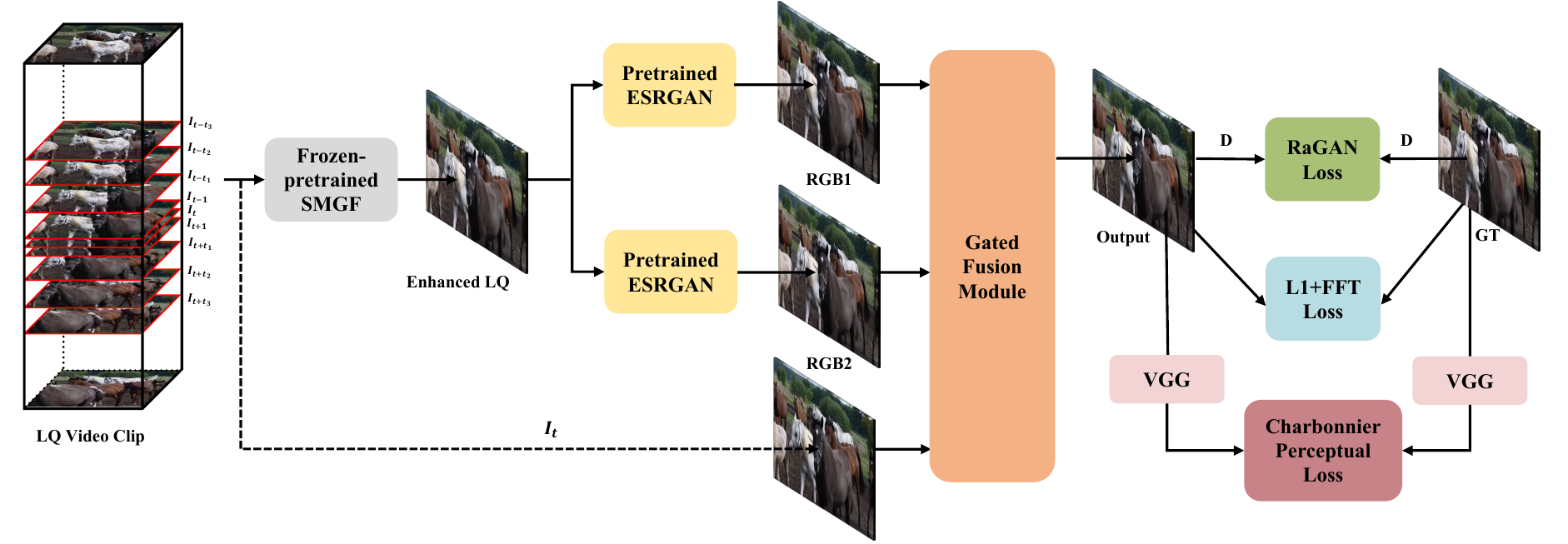}}
	\caption{Network architectures of the BILIBILI AI \& FDU Team. }
	\label{fig:fig1}
\end{figure}

\section{Challenge methods and teams}

\subsection{BILIBILI AI \& FDU Team}\label{bilibili}

\textbf{Track 1.} In Track 1, they propose a Spatiotemporal Model with Gated Fusion~(SMGF)~\cite{bilibili} for enhancing compressed video, based on \cite{deng2020spatio}. The pipeline of the proposed method is illustrated in the top of Figure~\ref{fig:fig1}-(a).

As the preliminary step, they first decode the bitstream to extract the the QP of each frame. Based on the QP value, they select 4 previous and 4 subsequent frames as the reference frames, so totally 9 frames~(including the target frame) are fed into the model. 1) Denoting the time stamp of the target frame as $t$, then both two adjacent frames~($t-1$ and $t+1$) are selected; 2) Then, they take the three previous Peak Quality Frames~(PQFs)~\cite{yang2018multi} and three subsequent PQFs as additional reference frames. 3) If there are no more reference frames and the number of selected reference frames is fewer than 4 in the previous part or the subsequent part, then they repeatedly pad it with the last selected frames until there are totally 8 reference frames.


\begin{figure*}[!t]
      \begin{center}
            \subfigure[An Overview of BasicVSR++]{\includegraphics[width=.54\linewidth]{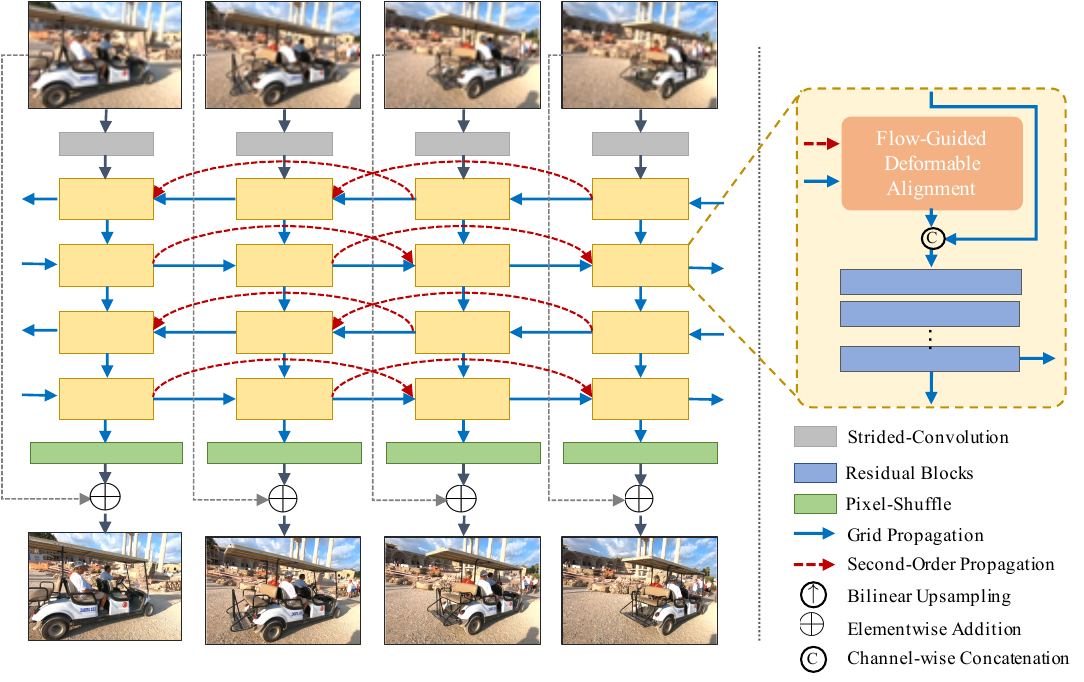}}
            \subfigure[Flow-guided deformable alignment.]{\includegraphics[width=.44\linewidth]{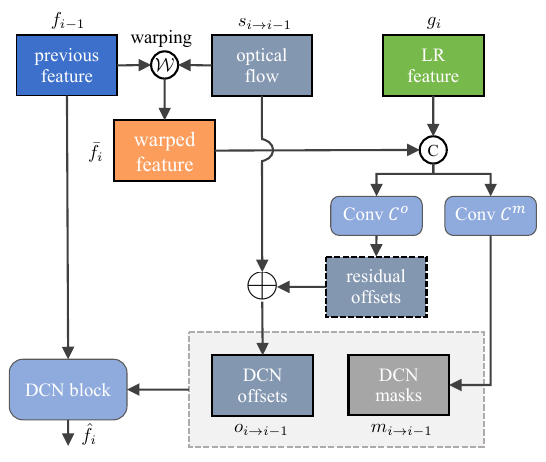}}
            \caption{The method proposed by the NTU-SLab Team.
            }
            \label{fig:overview}
      \end{center}
\end{figure*}

They feed 9 frames~(8 references and a target frame) into the Spaito-Temporal Deformable Fusion~(STDF)~\cite{deng2020spatio} module to capture spatiotemporal information.
The output of STDF module is then sent to the Quality Enchancement~(QE) module.
They employ a stack of adaptive WDSR-A-Block from C2CNet~\cite{fuoli2020ntire} as the QE module.
As illustrated in Figure~\ref{fig:fig1}, a Channel Attention~(CA) layer~\cite{zhang2018image} is additionally attached at the bottom of WDSR-A-Block~\cite{yu2018wide}.
Comparing with the CA layer in RCAN~\cite{zhang2018image}, there are two learnable parameters $\alpha$ and $\beta$ initialized with 1 and 0.2 in Ada-WDSR-A-Block.
Besides, the channels of the feature map and block in the QE module are 128 and 96, respectively. The channels of Ada-WDSR-A-Block are implemented as \{64, 256, 64\}.

Additionally, they propose a novel module to improve the performance of enhancement at the bottom of the pipeline.
As shown in the middle-top of Figure~\ref{fig:fig1}-(a), though each model has the same architecture~(STDF with QE) and training strategy~(L1 + FFT + Gradient~\cite{wang2020scene} loss), one is trained on the official training sets, and the other is on extra videos crawled from Bilibili~\cite{bili} and YouTube~\cite{youtube}, named as BiliTube4k.
To combine the predictions of two models, they exploit a stack of layers to output the mask $M$ and then aggregate predictions.
The mask $M$ in gated fusion module is with the same resolution of the target  frame ranging from $[0,1]$, the final enhanced low-quality frame is
formulated as
 \begin{equation}
 	\label{eq: gate-fusion}
	\hat{I}= M \otimes \hat{I}_1 \oplus (1-M)\otimes \hat{I}_2 .
\end{equation}

\textbf{Track 2.} In Track 2, they reuse and freeze the models of Track 1, and attach an ESRGAN~\cite{wang2018esrgan} at the bottom of SMGF to propose a perceptual SMGF~(pSMGF).  As shown in Figure~\ref{fig:fig1}-(b), they first take the enhanced low-quality frames from Track 1. Then they feed these enhanced frames into ESRGAN and train the Generator and Discriminator iteratively. Specifically, they use the ESRGAN pre-trained on DIV2K dataset~\cite{agustsson2017ntire}. They remove the pixel shuffle layer in ESRGAN, and supervise the model with \{L1 + FFT + RaGAN + Perceptual\} loss.
They also utilize the gated fusion module after ESRGAN, which is proposed in SMGF. Specifically, one of the ESRGANs is tuned on the official training sets, and the other is on extra videos collected from Bilibili~\cite{bili} and YouTube~\cite{youtube}, named as BiliTube4k. The predictions of two models are aggregated via \eqref{eq: gate-fusion}.

\textbf{Track 3.} They utilize the model in Track 1 as the pre-trained model, and then fine-tune it on the training data of Track 3 with early stopping.
Another difference is that they take the neighboring previous and subsequent I/P frames as reference frames, instead of PQFs.

\subsection{NTU-SLab Team}

\textbf{Overview.} The NTU-SLab Team proposes the BasicVSR++ method for this challenge. 
BasicVSR++ consists of two deliberate modifications for improving \textit{propagation} and \textit{alignment} designs of BasicVSR~\cite{chan2021basicvsr}. As shown in Figure~\ref{fig:overview}-(a), given an input video, residual blocks are first applied to extract features from each frame. The features are then propagated under the proposed second-order grid propagation scheme, where alignment is performed by the proposed flow-guided deformable alignment. After propagation, the aggregated features are used to generate the output image through convolution and pixel-shuffling.

\textbf{Second-order grid propagation.} Motivated by the effectiveness of the bidirectional propagation, they devise a grid propagation scheme to enable \textit{repeated refinement through propagation}. More specifically, the intermediate features are propagated backward and forward in time in an alternating manner. Through propagation, the information from different frames can be ``revisited'' and adopted for feature refinement. Compared to existing works that propagate features only once, grid propagation repeatedly extracts information from the entire sequence, improving feature expressiveness.
To further enhance the robustness of propagation, they relax the assumption of first-order Markov property in BasicVSR and adopt a second-order connection, realizing a second-order Markov chain.
With this relaxation, information can be aggregated from different spatiotemporal locations, improving robustness and effectiveness in occluded and fine regions.


\textbf{Flow-guided deformable alignment.} Deformable alignment~\cite{wang2019deformable,wang2019edvr} has demonstrated significant improvements over flow-based alignment~\cite{haris2019recurrent,xue2019video} thanks to the offset diversity~\cite{chan2021understanding} intrinsically introduced in deformable convolution (DCN)~\cite{dai2017deformable,zhu2019deformable}.
However, deformable alignment module can be difficult to train~\cite{chan2021understanding}. The training instability often results in offset overflow, deteriorating the final performance.
To take advantage of the offset diversity while overcoming the instability, they propose to employ optical flow to guide deformable alignment, motivated by the strong relation between deformable alignment and flow-based alignment~\cite{chan2021understanding}. The graphical illustration is shown in Figure~\ref{fig:overview}-(b).

\textbf{Training. }  The training consists of only one stage. For Tracks 1 and 3, only Charbonnier loss~\cite{charbonnier1994two} is used as the loss function. For Track 2, the perceptual and adversarial loss functions are also used. The training patch size is $256\times256$, randomly cropped from the original input images. They perform data augmentation, \ie, rotation ($0^{\circ},\, 90^{\circ},\, 180^{\circ},\, 270^{\circ}$), horizontal flip, and vertical flip. For Track 1, they initialize the model from a variant trained for video super-resolution to shorten the training time. The models for the other two tracks are initialized from the model of Track 1. During the test phase, they test the proposed models with ensemble ($\times 8$) testing, \ie, rotating $90^{\circ}$, flipping the input in four ways (none, horizontally, vertically, both horizontally and vertically) and averaging their outputs.

\subsection{VUE Team}

\begin{figure}[!t]
\centering
\subfigure[The method for Track 1.]{\includegraphics[width=.8\linewidth]{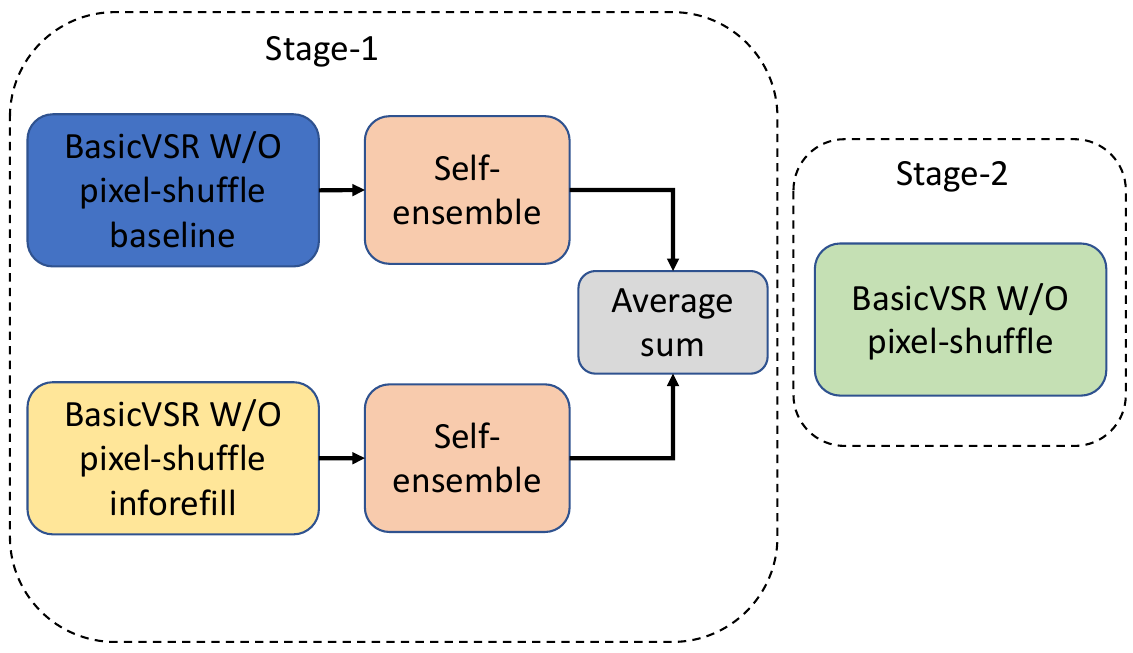}}
\subfigure[The method for Track 3.]{\includegraphics[width=.5\linewidth]{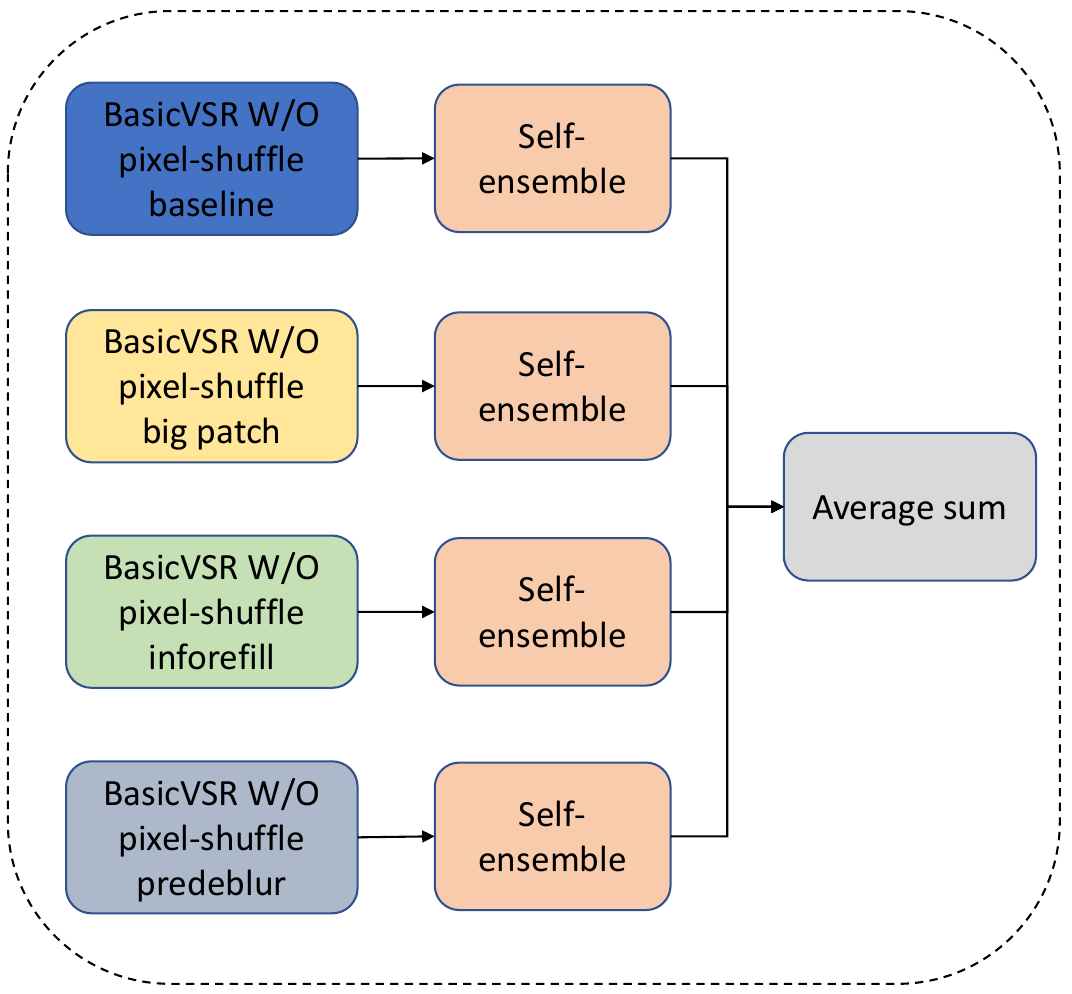}}
\subfigure[BasicVSR w/o pixel-shuffle.]{\includegraphics[width=.45\linewidth]{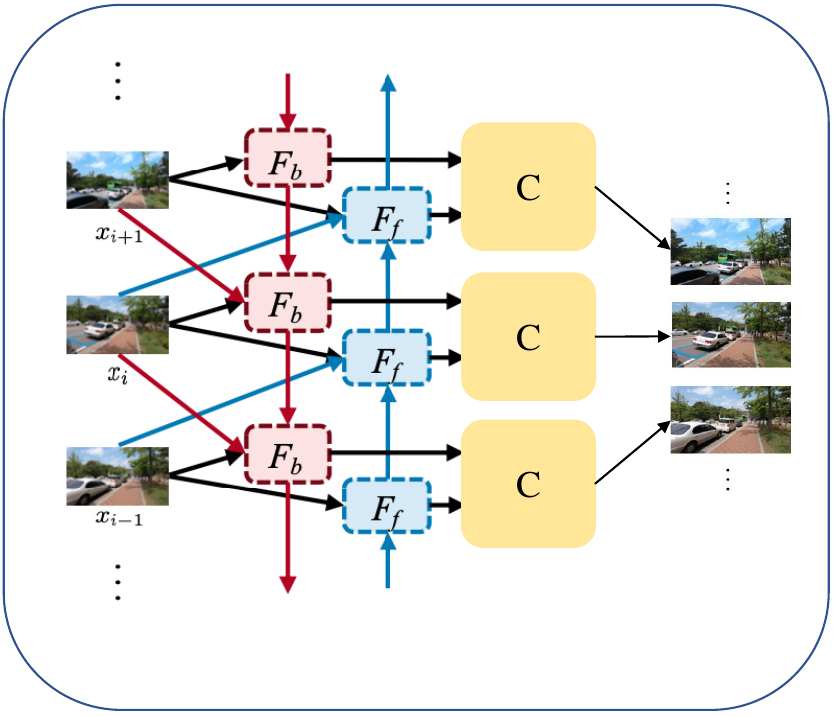}}
\caption{The methods of the VUE Team for Tracks 1 and 3.}
\label{fig:VUE}
\end{figure}

\textbf{Tracks 1 and 3.} In the fidelity tracks, the VUE Team proposes the methods based on BasicVSR~\cite{chan2021basicvsr}, as shown in Figure~\ref{fig:VUE}. For Track 1, they propose a two-stage method. In stage-1, they train two BasicVSR models with different parameters followed by the self-ensemble strategy. Then, they fuse the two results by calculating the average of them. In stage-2, they train another BasicVSR model. For Track 3, they propose to tackle this problem by using VSR methods without the last upsampling layer. They train four BasicVSR models with different parameter settings followed by the self-ensemble strategy. Then, they average the four outputs as the final result.

\begin{figure}[!t]
\centering
\includegraphics[width=1\linewidth]{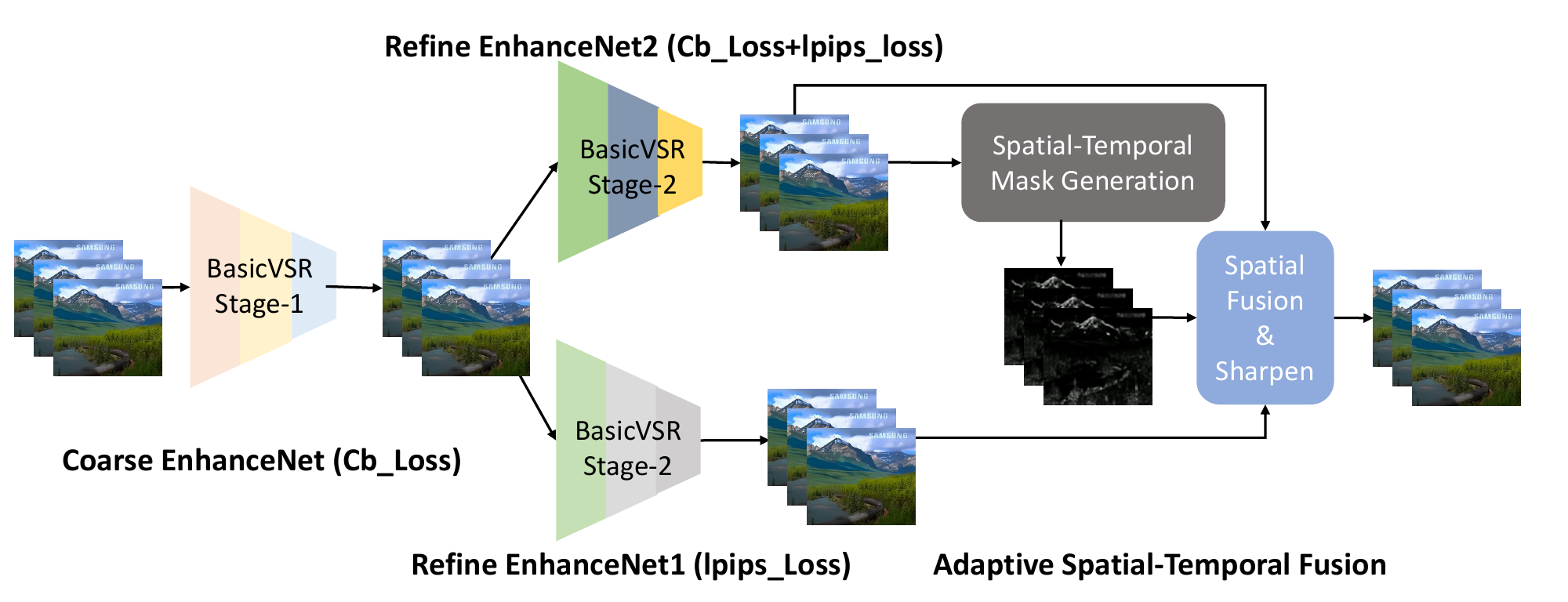}
\caption{The proposed method of the VUE Team for Track 2.}
\label{fig:VUE_2}
\end{figure}

\textbf{Track 2. } In Track 2, they propose a novel solution dubbed ``Adaptive Spatial-Temporal Fusion of Two-Stage Multi-Objective Networks''~\cite{li2021VUE}. It is motivated by the fact that it is hard to design unified training objectives which are perceptual-friendly for enhancing regions with smooth content and regions with rich textures simultaneously.
To this end, they propose to adaptively fuse the enhancement results from the networks trained with two different optimization objectives. As shown in Figure~\ref{fig:VUE_2}, the framework is designed with two stages. The first stage aims at obtaining the relatively good intermediate results with high fidelity. In this stage, a BasicVSR model is trained with Charbonnier loss~\cite{charbonnier1994two}. At the second stage, they train two BasicVSR models for different refinement purposes. One refined BasicVSR model (denoted as EnhanceNet2) is trained with 
\begin{equation}
3\cdot \text{Charbonnier loss} + \text{LPIPS loss}.
\end{equation}
Another refined BasicVSR model (denoted as EnhanceNet1) is trained with the mere LPIPS loss~\cite{zhang2018unreasonable}. This way, EnhanceNet1 is good at recovering textures to satisfying human perception requirement but it can result in temporal flickering for smooth regions of videos, meanwhile EnhanceNet1 produces much more smooth results, especially, temporal flickering is well eliminated.

To overcome this issue, they devise a novel adaptive spatial-temporal fusion scheme. Specifically, the spatial-temporal mask generation module is proposed to produce spatial-temporal mask and it is used to fuse the outputs of the two networks:
\begin{equation}
    I_{out}^t = (1 - mask_{t}) \times I_{out, 1}^{t} + mask_{t} \times I_{out,2}^{t},
\end{equation}
where $mask_t$ is the generated mask for the $t$-th frame, $I_{out,1}^t$ and $I_{out,2}^t$ are the $t$-th output frames of EnhanceNet1 and EnhanceNet2, respectively. The mask $mask_t=f(I_{out,2}^{t-1},I_{out,2}^{t},I_{out,2}^{t+1})$ is adaptively generated from $I_{out,2}^{t-1}$, $I_{out,2}^{t}$ and $I_{out,2}^{t+1}$ as follows. First, the variance map $V^{t}$ is calculated from $I_{out,2}^{t}$ by:
\begin{equation}
    \begin{split}
    &V^{t}_{i,j} = Var(Y^{t}_{out,2}[i-5:i+5,j-5:j+5]) \\
    &Y^{t}_{out,2} = (I^{t}_{out,2}[:,:,0] + I^{t}_{out,2}[:,:,1] + I^{t}_{out,2}[:,:,2])/3,
    \end{split}
\end{equation}
where $Var(x)$ means the variance of $x$. Then, they normalize the variance map in a temporal sliding window to generate the mask $mask_t$:
\begin{equation}
\begin{split}
    &mask_t = (V^t-q)/(p-q) \\
    &p = max([V^{t-1}, V^t, V^{t+1}]) \\
    &q = min([V^{t-1}, V^t, V^{t+1}]).
\end{split}
\end{equation}
Intuitively, when a region is smooth, its local variance is small, otherwise, its local variance is large. Therefore, smooth region more relies on the output of EnhanceNet2 while the rich-texture region gets more recovered details from EnhanceNet1. With temporal sliding window, the temporal flickering effect is also well eliminated.

\subsection{NOAHTCV Team}

\begin{figure}[!t]
\centering
\includegraphics[width=1\linewidth]{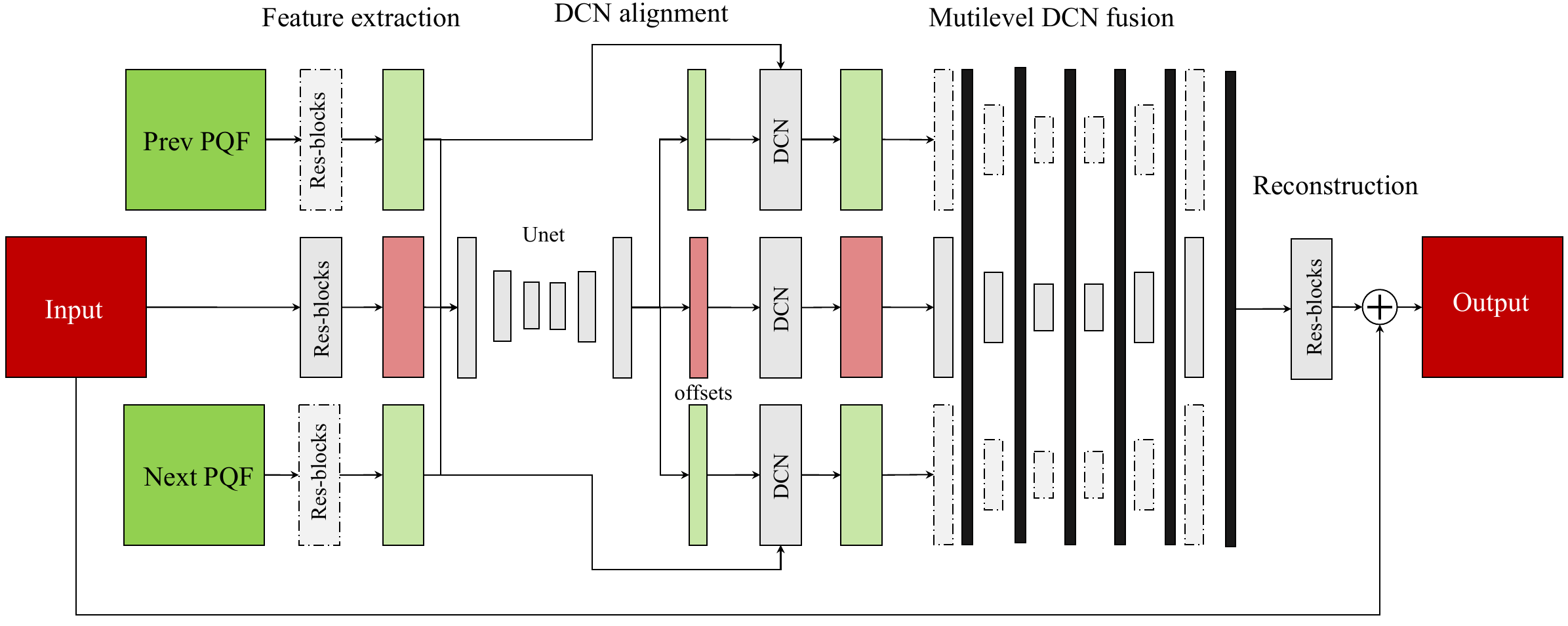}
\caption{The proposed method of the NOAHTCV Team.}
\label{fig:model}
\end{figure}

As show in Figure~\ref{fig:model}, the input images include three frames, \ie, the current frame plus the previous and the next Peak Quality Frames (PQF). The first step consists of a shared feature extraction with a stack of residual blocks and subsequently an U-Net is used to jointly predict the individual offsets for each of the three input. Such offsets are then used to implicitly align and fuse the features. Note that, there is no loss used as supervision for this step. After the initial feature extraction and alignment, they use a multi-head U-Net with shared weights to process each input feature, and at each scale of the encoder and decoder, they fuse the U-Net features with scale-dependant deformable convolutions, which are shown in black in Figure~\ref{fig:model}. The output features of the U-Net are fused for a final time, and the output fused features are finally processed by a stack of residual blocks to predict the final output. This output is in fact residual information which is added to the input frame to produce the enhanced output frame. The models utilized for all three tracks are the same. The difference is the loss function, \ie, they use the $L2$ loss for Tracks 1 and 3, and use GAN Loss + Perceptual loss + $L2$ loss for Track 2.

\subsection{MT.MaxClear Team}

The proposed model is based on EDVR~\cite{wang2019edvr}, which uses the deformable convolution to align features between neighboring frames and the target frame, and then combines all aligned frame features to reconstruct the target frame.  The deformable convolution module in EDVR is difficult to train due to the unstable of DCN offset. They propose to add two DCN offset losses to regularize the deformable convolution module which makes the training of DCN offset much more stable. They use Charbonnier penalty loss~\cite{charbonnier1994two}, DCN offsets Total Variation loss and DCN offsets Variation loss to train the model. The Charbonnier penalty loss is more robust than $L2$ loss. DCN offsets Total Variation loss encourages the predicted DCN offsets  are smooth in spatial space. DCN offsets Variation loss encourages the predicted DCN offsets between different channels do not deviate too much from the offsets mean. The training of DCN is much more stable due to the aforementioned two offsets losses, but the EDVR model performs better if the loss weights of DCN offsets Total Variation loss and DCN offsets Variation loss gradually decays to zero during training. In Track 2, they add the sharpening operation on the enhanced frames for better visual perception.

\subsection{Shannon Team}

The Shannon Team introduces a disentangled attention for compression artifact analysis. Unlike
previous works, they propose to address the problem of artifact reduction from a
new perspective: disentangle complex artifacts by a disentangled attention mechanism.
Specifically, they  adopt a multi-stage architecture in which early stage also provides
a disentangled attention. Their key insight is that there are various types
of artifacts created by video compression, some of which result in significant
blurring effect in the reconstructed signals, and some of which generate artifacts, such as blocking and ringing.  Algorithms may be either too aggressive and amplify erroneous high-frequency
components, or be too conservative and tend to smooth over ambiguous components.
Both result in bad cases that seriously affect subjective visual impression.
The proposed disentangled attention aims to reduce these bad cases.

\begin{figure}[!t]
	\centering
	\includegraphics[width=1\linewidth]{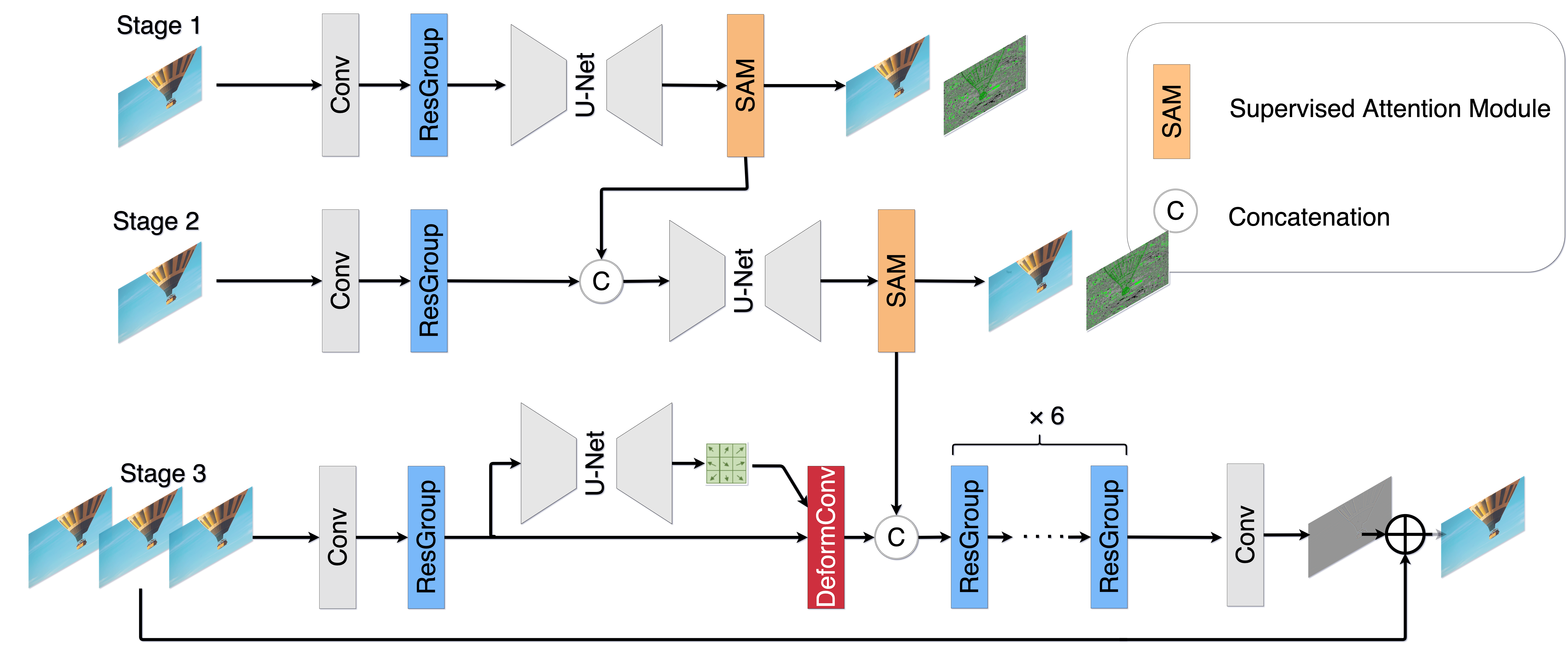}
	\caption{The proposed generator of the Shannon Team. 
		}
	\label{fig:shannon}
\end{figure}

In Track 2, they use the LPIPS loss and only feed the high-frequency components to the discriminator.
Before training the model, they analyze the quality fluctuation among frames \cite{yang2018multi}
and train the model from-easy-to-hard. To generate the attention map, they use
the supervised attention module proposed in \cite{mehri2021mprnet}.
The overall structure of the proposed generator is shown in Figure~\ref{fig:shannon}. The discriminator
is simply composed of several convolutional layer-ReLU-strided convolutional
layers blocks, and its final output is a 4×4 confidence map.

Let $F_{lp}$ denote the low-pass filtering, which is implemented in a differentiable manner with \textit{Kornia} \cite{riba2020kornia}. They derive the supervised information for attention map:
\begin{equation} 
	\mathcal{R}_{lp} = F_{lp}(x) - x,
	\label{eq:a2}
\end{equation}
\begin{equation} 
	\mathcal{R}_{y} = y - x, 
	\label{eq:a1}
\end{equation}
\begin{equation} 
	D(y, x) = sgn(\mathcal{R}_{y} \odot \mathcal{R}_{lp}),
	\label{eq:a3}
\end{equation}
where $sgn(\cdot)$ denotes the signum function that extracts the sign of a given pixel value; $\odot$ is the element-wise product, and $y$ refers to the output, and $x$ refers to the compressed input.

\subsection{Block2Rock Noah-Hisilicon Team}

\begin{figure}[!t]
\centering
\includegraphics[width=1\linewidth]{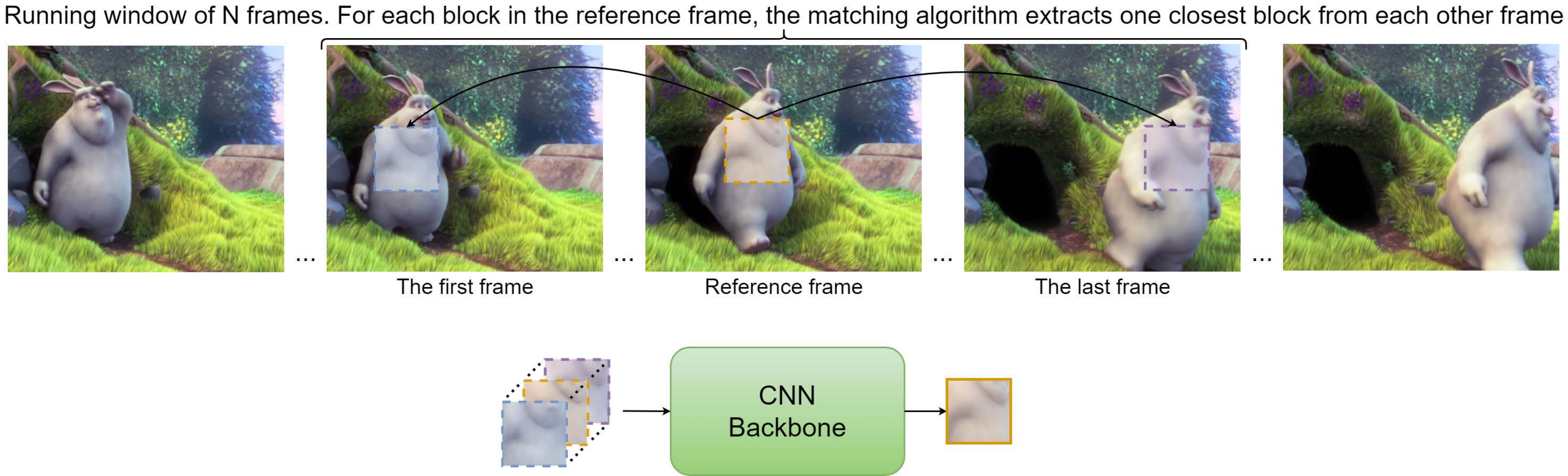}
\caption{The illustration of the proposed method of the Block2Rock Noah-Hisilicon Team.}
\label{fig:Block2Rock}
\end{figure}

This team makes a trade-off between spatial and temporal sizes in favor of the latter by performing collaborative CNN-based restoration of square patches extracted by block matching algorithm, which finds correlated areas across consequent frames. Due to performance concerns, the proposed block matching realization is trivial: for each patch $\vect{p}$ at the reference frame, they search for and extract a single closest patch $\vect{p}_i$ from each other frame $\vect{f}_i$ in a sequence, based on squared $L2$ distance:
\begin{equation}\label{eq:patch_search}
    \vect{p}_i = \vect{C}(\hat{u}, \hat{v}) \vect{f}_i, \ \textrm{where} \ \hat{u}, \hat{v} = \argmin_{u, v} || \vect{C}(u, v) \vect{f}_i - \vect{p} ||^2_2.
\end{equation}
Here $\vect{C}(u, v)$ is a linear operator that crops patch whose top-left corner is located at $(u, v)$ pixel coordinates of the canvas. As it is shown for example in \cite{google_hdr}, the search for the closest patch in \eqref{eq:patch_search} requires a few pointwise operations and two convolutions (one of which is a box filter), which can be done efficiently in the frequency domain based on convolution theorem. The resulted patches are then stacked and passed to a CNN backbone, which outputs a single enhanced version of the reference patch. The overall process is presented in Figure~\ref{fig:Block2Rock}.

Since the total number of pixels being processed by CNN depends quadratically on spatial size and linearly on the input's temporal size, they propose to use inference on small patches of size $48 \times 48$ pixels to decrease the spatial size of backbone inputs. For example, the two times decrease in height and width allows increasing temporal dimension by a factor of four. With existing well-performing CNNs, this fact allows the temporal dimension to increase up to $50$ frames since such models are designed to be trained on patches with spatial sizes of more than $100$ pixels (typically 128 pixels).

In the solution, they use the EDVR network~\cite{wang2019edvr} as a backbone, and the RRDB network \cite{rrdb} acts as a baseline. For EDVR, they stack patches in a separate dimension, while for RRDB, they stack patches in channel dimension. Reference patch is always the first in a stack.

For training network weights, they use the $L1$ distance between output and target as an objective to minimize through back propagation and stochastic gradient descent. They use the Adam optimizer~\cite{kingma2014adam} with learning rate increased from zero to $2e^{-4}$ during a warmup period, which then is gradually decreased by a factor of $0.98$ after each epoch. The total number of epochs was $100$ with $2000$ unique batches passed to the network during each one. To stabilize the training and prevent divergence, they use the adaptive gradient clipping technique with weight $\lambda = 0.01$, as proposed in~\cite{agc}.

\subsection{Gogoing Team}

\begin{figure}
\centering
\subfigure[Overall architecture]{\includegraphics[width=1\linewidth]{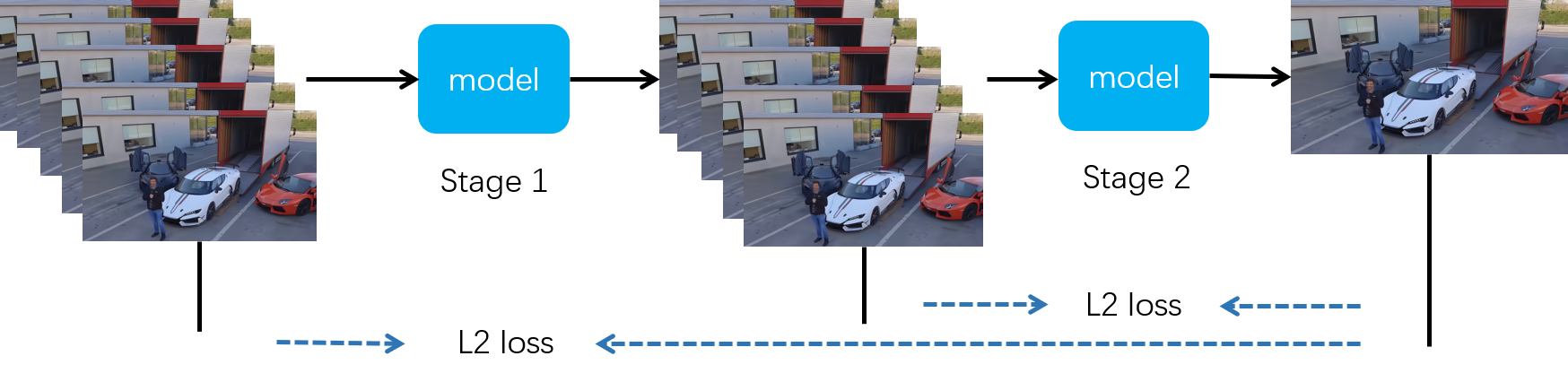}}
\subfigure[Model architecture]{\includegraphics[width=1\linewidth]{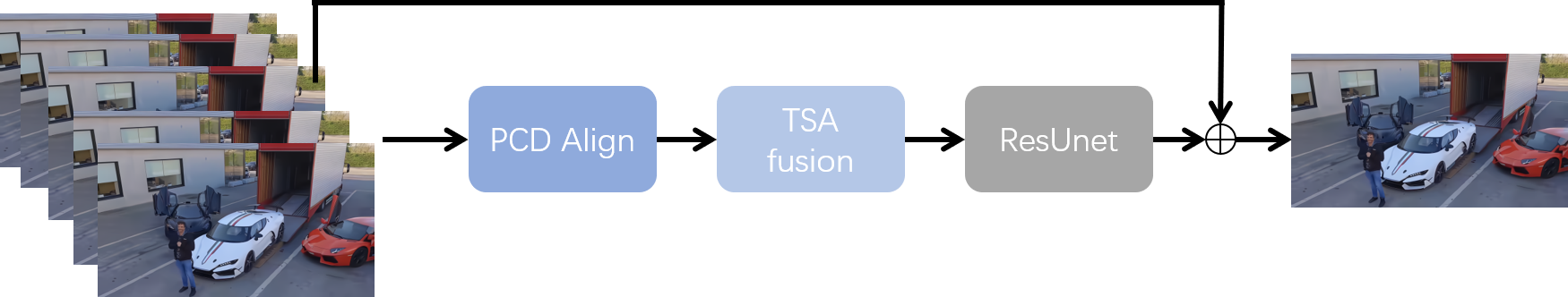}}
\subfigure[ResUNet architecture]{\includegraphics[width=1\linewidth]{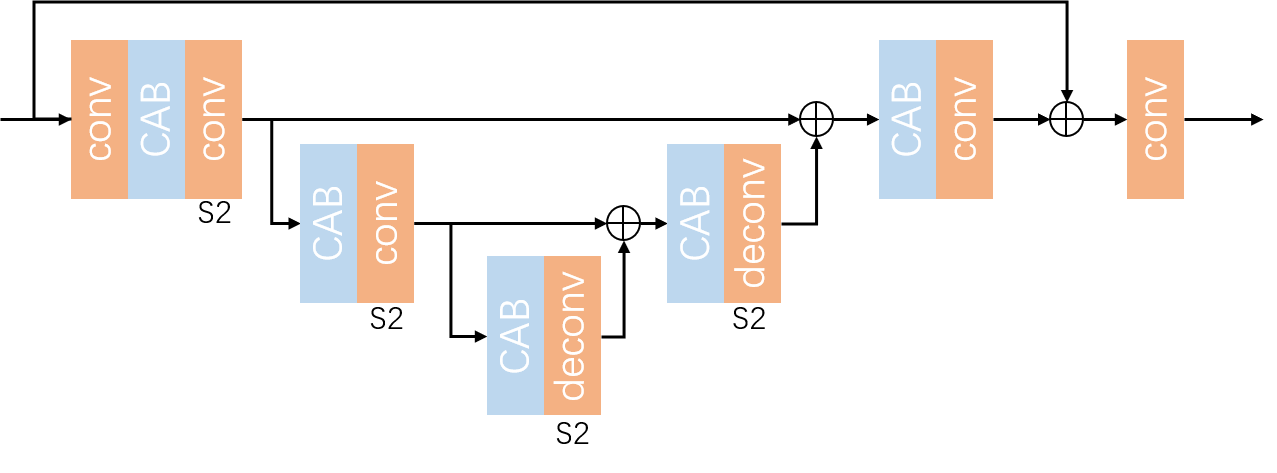}}
\caption{The proposed method of Gogoing Team.}
\label{fig:gogoing}
\end{figure}

The overall structure adopts a two-stage model, as shown in the Figure~\ref{fig:gogoing}-(a). As Figure~\ref{fig:gogoing}-(b) shows, for the temporal part, they use the temporal module in EDVR~\cite{wang2019edvr}, that contains the PCD module and the TSA module. The number of input frames is 7. In the spatial part, they combine the UNet~\cite{nah2020ntire} and the residual attention module~\cite{zhang2018image} to form a ResUNet, as shown in Figure~\ref{fig:gogoing}-(c). In the training phase, they use the 256$\times$256 RGB patchs from the training set as input, and augment them with random horizontal flips and $90^{\circ}$ rotations. All models are optimized by using the Adam~\cite{kingma2014adam} optimizer with mini-batches of size 12, with the learning rate being initialized to $4\times10^{-4}$ using the CosineAnnealingRestartLR strategy. Loss function is the $L2$ loss.

\subsection{NJU-Vision Team}

\begin{figure}[!t]
\centering
\includegraphics[width=1\linewidth]{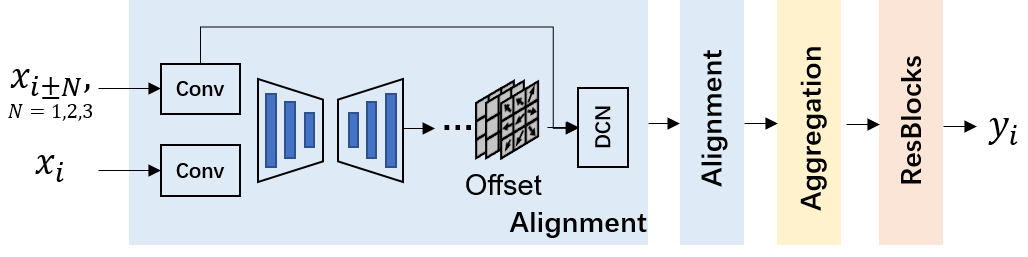}
\caption{The proposed method of the NJU-Vision Team.}
\label{fig:NJU-Vision}
\end{figure}

As shown in Figure~\ref{fig:NJU-Vision}, the NJU-Vision Team proposes a method utilizing a progressive deformable alignment module and a spatial-temporal attention based aggregation module, based on~\cite{wang2019edvr}. Data augmentation is also applied with training data augmentation by randomly flipping in horizontal and vertical orientations, and rotating at $90^{\circ}$, $180^{\circ}$, and $270^{\circ}$, and evaluation ensemble by flipping in horizontal and vertical orientations, and rotating at $90^{\circ}$, $180^{\circ}$, and $270^{\circ}$ to obtain the averaged results.

\begin{figure}[!t]
	\centering
	\subfigure[3D–MultiGrid BackProjection network (MGBP–3D).]{\includegraphics[width=\linewidth]{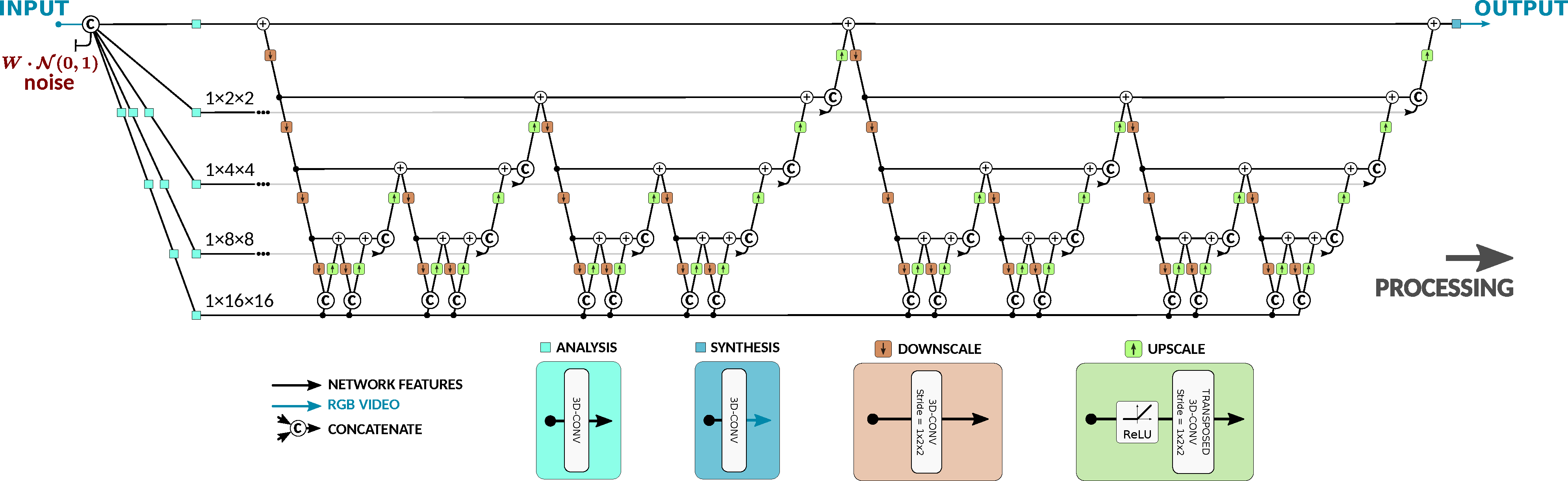}}\\
	\subfigure[Multiscale discriminator.]{\includegraphics[width=\linewidth]{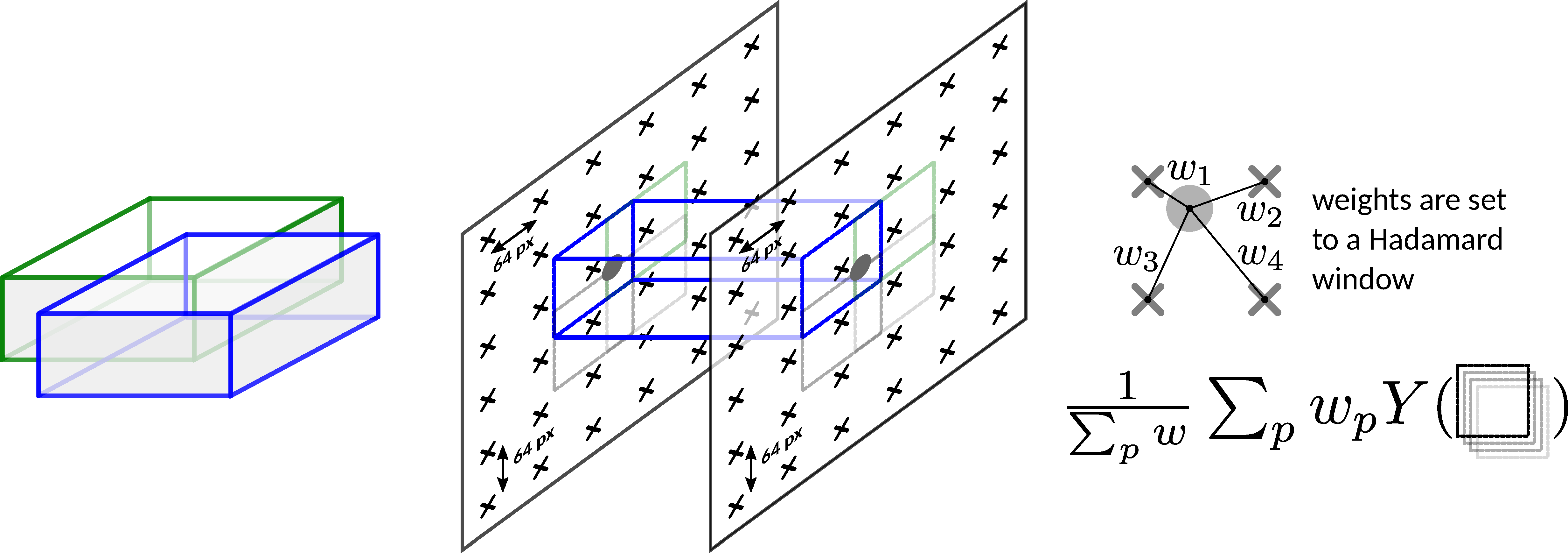}}\\
		\subfigure[Overlapping–patches inference strategy.]{\includegraphics[width=\linewidth]{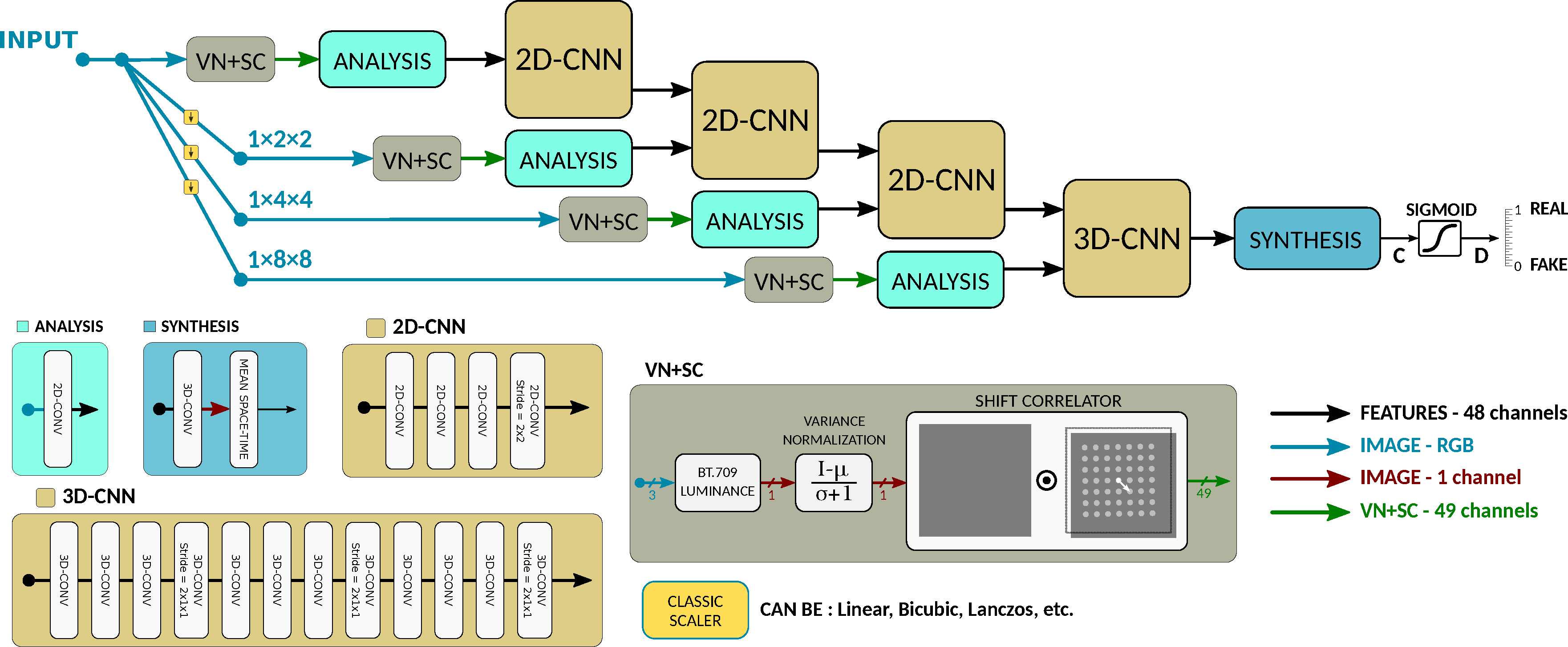}}
	\caption{Network architectures of the BOE-IOT-AIBD Team. }
	\label{fig:BOE}
\end{figure}

\subsection{BOE-IOT-AIBD Team}

\textbf{Tracks 1 and 3. } Figure~\ref{fig:BOE}-(a) displays the diagram of the MGBP–3D network used in this challenge, which was proposed by the team members in \cite{michelini2021multi}. The system uses two backprojection residual blocks that
run recursively in five levels. Each level downsamples only space, not
time, by the factor of 2. The Analysis and Synthesis modules convert an image
into features space and vice–versa using single 3D–convolutional layers.
The Upscaler and Downscaler modules are composed of single strided
(transposed and conventional) 3D–convolutional layers. Every Upscaler
and Downscaler module shares the same configuration in a given level but
they do not share parameters. Small number of features are set at high
resolution and they increase at lower resolutions to reduce the memory
footprint in high resolution scales. In addition, they add a 1–channel noise
video stream to the input that is only used for the Perceptual track. In the fidelity tracks, the proposed model is trained by the $L2$ loss.

To process long video sequences they use the patch based approach from \cite{michelini2021multi}, in which they average the output of overlapping video patches taken
from the compressed degraded input. First, they divide input streams into
overlapping patches (of same size as training patches) as shown in Figure~\ref{fig:BOE}-(b); second, they multiply each output by the weights set to a Hadamard window;
and third, they average the results. In the experiments they use overlapping
patches separated by 243 pixels in vertical and horizontal directions and one
frames in time direction. 

\textbf{Track 2. } Based on the model for Tracks 1 and 3, they add noise inputs to activate and deactivate the generation of artificial
details for the Perceptual track. In MGBP–3D, they generate one channel
of Gaussian noise concatenated to the bicubic upscaled input. The noise
then moves to different scales in the Analysis blocks. This change allows
using the overlapping patch solution with noise inputs, as it simply
represent an additional channel in the input. 

They further employ a discriminator shown in Figure~\ref{fig:BOE}-(c) to achieve adversarial training. The loss function used for Track 2 is a combination of the GAN loss, LPIPS loss and the $L1$ loss. $Y_{n=0}$ and $Y_{n=1}$ are the outputs of the generator using noise amplitudes $n=0$ and $n=1$, respectively. $X$ indicates the groundtruth. The loss function can be expressed as as follows:
\begin{equation}
 \begin{aligned}
    \mathcal{L}(Y, X; \theta) = \; & 0.001 \cdot \mathcal{L}^{RSGAN}_G(Y_{n=1}) \;  \\
     & + 0.1 \cdot \mathcal{L}^{perceptual}(Y_{n=1}, X) \; \\
     & + 10 \cdot \mathcal{L}^{L1}(Y_{n=0}, X) \;,
\end{aligned}
\end{equation}
Here, $\mathcal{L}^{perceptual}(Y_{n=1}, X)=\text{LPIPS}(Y_{n=1}, X)$ and the Relativistic GAN loss~\cite{jolicoeur2018relativistic}, is given by:
\begin{equation}
\begin{aligned}
&\mathcal{L}_D^{RSGAN} 
\\&= -\mathbb{E}_{(real,fake) \sim (\mathbb{P},\mathbb{Q})}\left[ \log (\text{sigmoid}(C(real)-C(fake))) \right], \nonumber
\end{aligned}
\end{equation}
\begin{equation}\label{12}
\begin{aligned}
&\mathcal{L}_G^{RSGAN}\\
&=  -\mathbb{E}_{(real,fake) \sim (\mathbb{P},\mathbb{Q})}\left[ \log (\text{sigmoid}(C(fake)-C(real))) \right],
\end{aligned}
\end{equation}
where $C$ is the output of the discriminator just before the sigmoid function, as shown in Figure~\ref{fig:BOE}-(c). In \eqref{12}, $real$ and $fake$ are the sets of inputs to the discriminator, which contains multiple inputs with multiple scales, \ie,
\begin{equation}
fake = \left\{ Y_{n=1} \right\},\ \ real = \left\{ X \right\}.
\end{equation}
After every epoch they evaluate the current model according to the validation metric:
\begin{align}
    \mathcal{V}(Y; \theta) = \mathbb{E}\Big[ & \text{LPIPS}(Y_{n=1}, X)\Big] \;.
\end{align}
The motivation is a simple and automatic metric that can help to find a model that generates realistic images in the full resolution.

\subsection{(\textit{anonymous})}

\begin{figure}
\centering
\includegraphics[width=.9\linewidth]{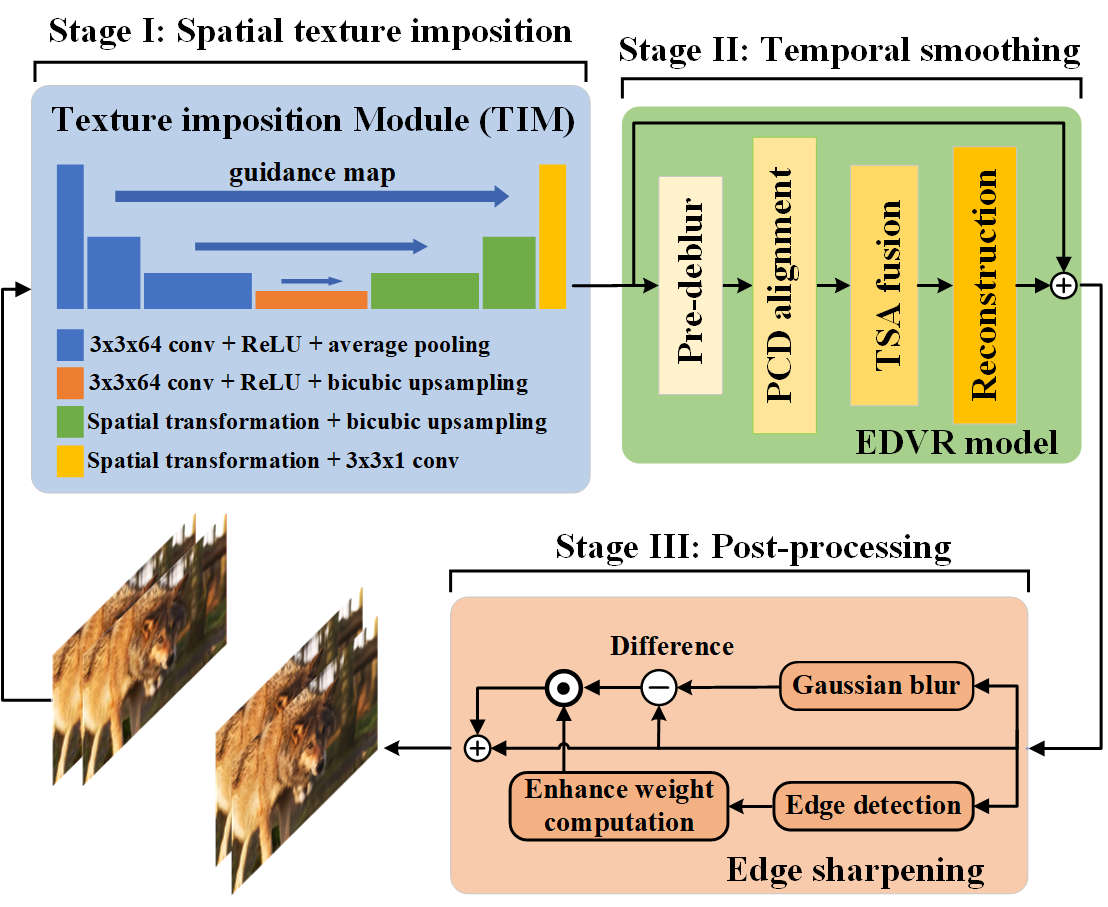}
\caption{The proposed method of (\textit{anonymous}).}
\label{fig:architecture}
\vspace{-1em}
\end{figure}

\textbf{Network. } For enhancing the perceptual quality of heavily compressed videos, there are two main problems that need to be solved: spatial texture enhancement and temporal smoothing.
Accordingly, in Track 2, they propose a multi-stage approach with specific designs for the above problems.
Figure~\ref{fig:architecture} depicts the overall framework of the proposed method, which consists of three processing stages.
In stage I, they enhance the distorted textures of manifold objects in each compressed frame by the proposed Texture Imposition Module (TIM).
In stage II, they suppress the flickering and discontinuity of the enhanced consecutive frames by a video alignment and enhancement network, \ie, EDVR~\cite{wang2019edvr}.
In stage III, they further enhance the sharpness of the enhanced videos by several classical techniques (instead of learning-based network).

In particular, TIM is based on the U-Net~\cite{unet} architecture to leverage the multi-level semantic guidance for texture imposition.
In TIM, natural textures of different objects in compressed videos are assumed as different translation styles, which need to be learned and imposed;
thus, they apply the affine transformations~\cite{stf} in the decoder path of the U-Net, to impose the various styles in a spatial way.
The parameters of the affine transformations are learned from several convolutions with the input of guidance map from the encoder path of the U-Net.
Stage II is based on the video deblurring model of EDVR, which consists of four modules, the PreDeblure, the PCD Align, the TSA fusion and the reconstruction module. In stage III, a combination of classical unsharp masking~\cite{deng2010generalized} and edge detection~\cite{sobel1972camera} methods are adopted to further enhance the sharpness of the video frames. In particular, they first obtain the difference map between the original and its Gaussian blurred images. Then, they utilize the Sobel operator~\cite{sobel1972camera} to detect the edges of original image to weight the difference map, and add the difference map to the original image.

\begin{figure*}[!t]
\centering
\subfigure[The overall architecture of the CVQENet.]{\includegraphics[width=.5\linewidth]{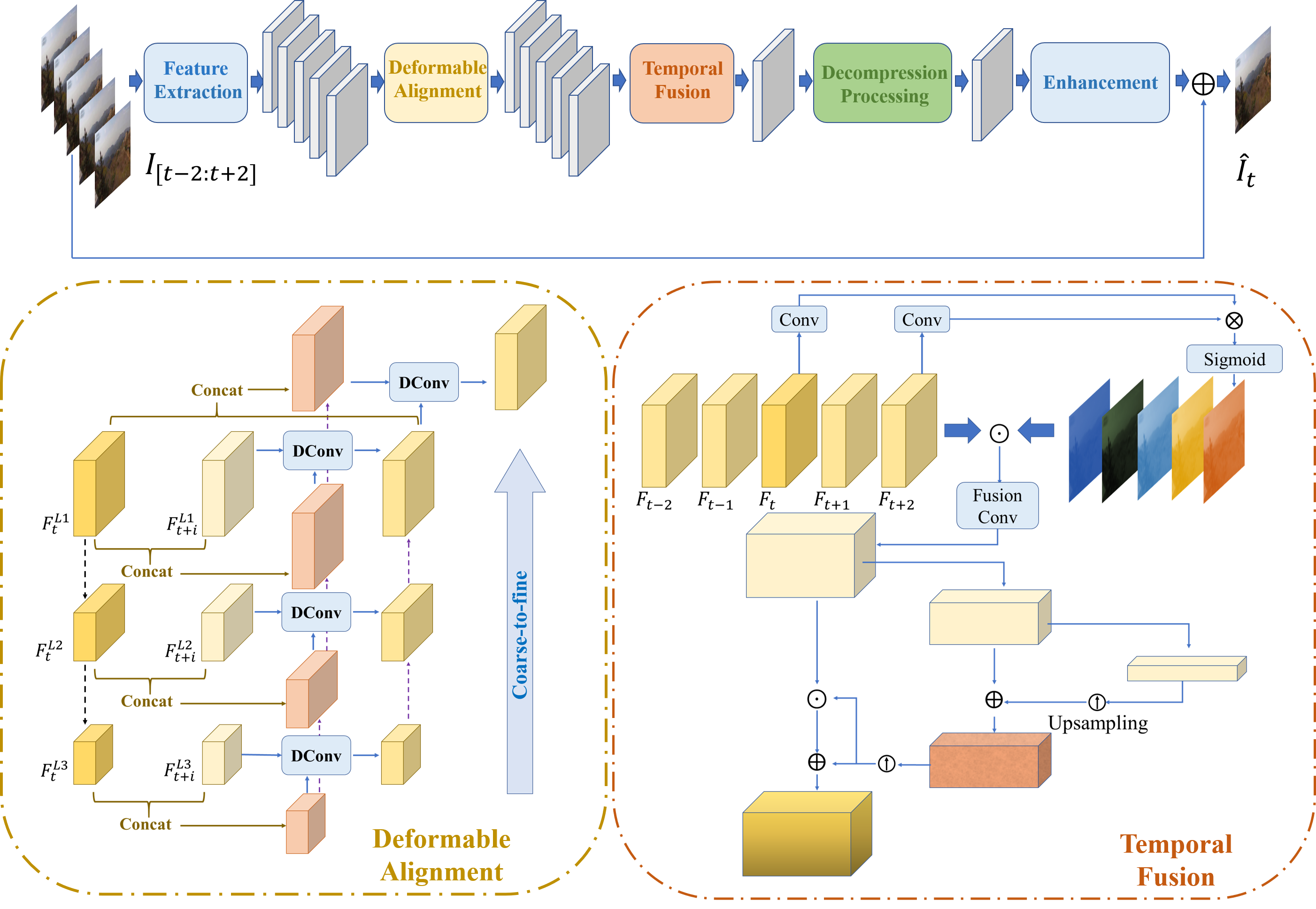}}
\hspace{2em}
\subfigure[The architecture of the DPM.]{\includegraphics[width=.4\linewidth]{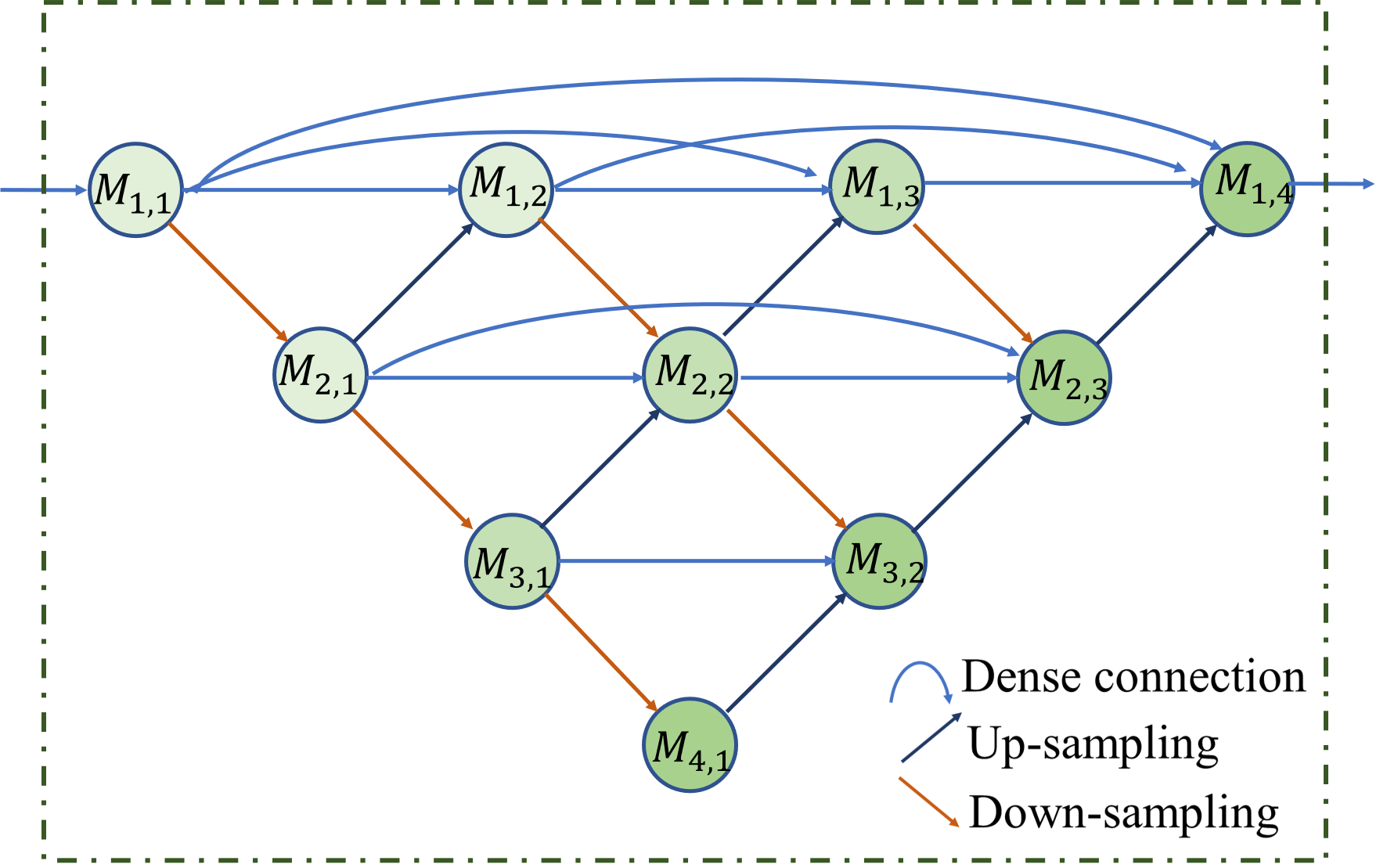}}
\caption{The proposed method of the VIP\&DJI Team for Track 1.}
\label{fig:vip}
\end{figure*}

\begin{figure*}[!t]
\centering
\subfigure[The whole framework of the DUVE.]{\includegraphics[width=.6\linewidth]{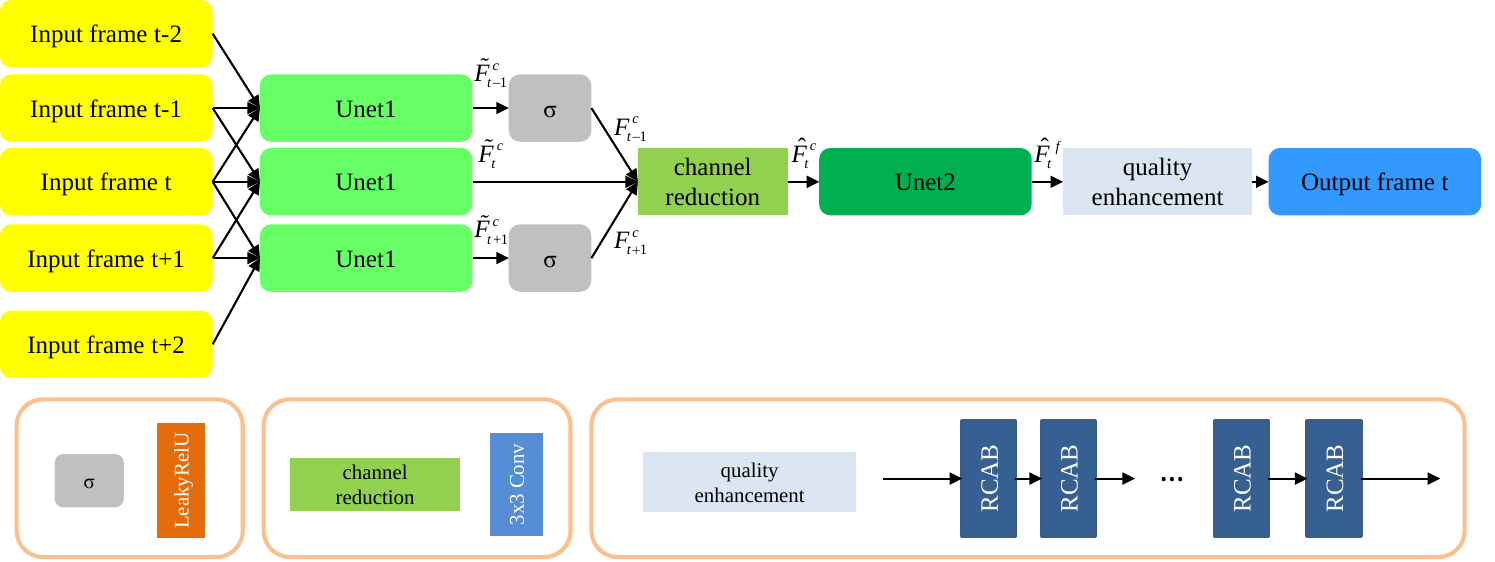}}
\hspace{1em}
\subfigure[The architecture of U-Net.]{\includegraphics[width=.35\linewidth]{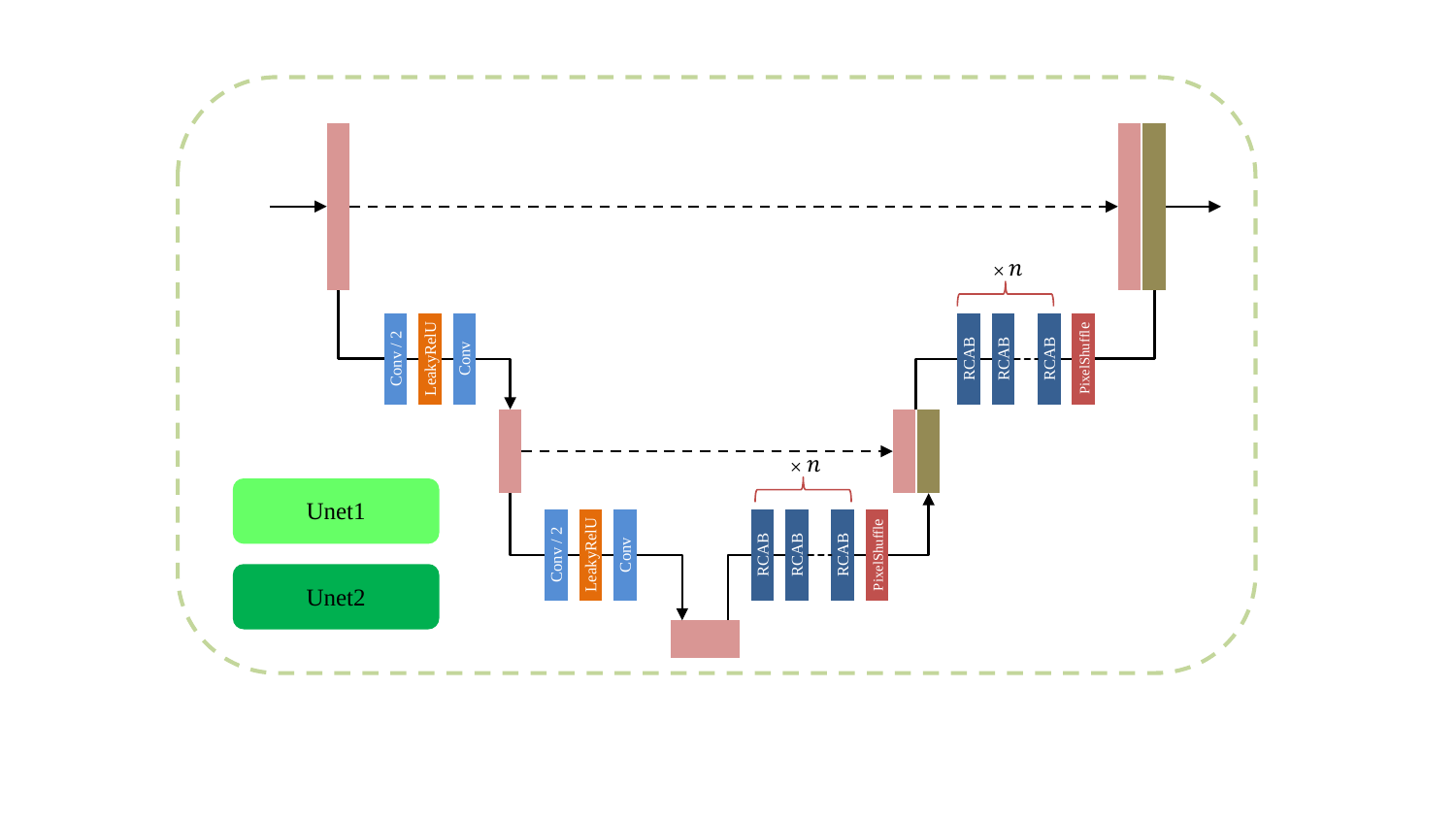}}
\caption{The proposed method of the VIP\&DJI Team for Track 3.}
\label{fig:vip2}
\end{figure*}

\textbf{Training. } 
For stage I, TIM is supervised by three losses: Charbonnier loss (epsilon is set to be $1e{-6}$) $\mathcal{L}_{\text{pixel}}$, VGG loss $\mathcal{L}_{\text{vgg}}$ and Relativistic GAN loss $\mathcal{L}_{\text{gan}}$.
The overall loss function is defined as: $\mathcal{L} = 0.01 \times \mathcal{L}_{\text{pixel}} + \mathcal{L}_{\text{gan}} + 0.005 \times \mathcal{L}_{\text{vgg}}$.
For stage II, they fine-tune the video deblurring model of EDVR with the training set of NTIRE, supervised by the default Charbonnier loss.
Instead of the default training setting in~\cite{wang2019edvr}, they first fine-tune the PreDeblur module with 80,000 iterations.
Then they fine-tune the overall model with a small learning rate of $1\times10^{-5}$ for another 155,000 iterations.

\subsection{VIP\&DJI Team}

\subsubsection{Track 1}

As shown in Figure~\ref{fig:vip}-(a), the architecture of the proposed CVQENet consists of five parts, which are feature extraction module, inter-frame feature deformable alignment module, inter-frame feature temporal fusion module, decompression processing module and feature enhancement module.
The input of CVQENet includes five consecutive compressed video frames $I_{[t-2:t+2]}$, and the output is a restored middle frame $\hat{I}_t$ that is as close as possible to the uncompressed middle frame $O_t$:
\begin{equation}
\hat{I}_t = \text{CVQENet}(I_{[t-2:t+2]}, \theta)
\end{equation}
where $\theta$ represents the set of all parameters of the CVQENet.
Given a training dataset, the loss function $L$ is defined as below:
\begin{equation}
L = || \hat{I}_t - O_t ||_1
\end{equation}
where $||.||_1$ denotes the $L1$ loss.
The modules in CVQENet are to be introduced in details in the following.

\textbf{Feature Extraction Module (FEM).} The feature extraction module contains one convolutional layer (Conv) to extract the shallow feature maps $F^0_{[t-2:t+2]}$ from the compressed video frames $I_{[t-2:t+2]}$ and 10 stacked residual blocks without Batch Normalization (BN) layer to further process the feature maps.

\textbf{Inter-frame Feature Deformable Alignment Module (FDAM).}
Next, for the feature maps extracted by the FEM, FDAM aligns the feature map corresponding to each frame to the middle frame that needs to be restored.
It can be aligned based on optical flow, deformable convolution, 3D convolution, and other methods, and CVQENet uses the method based on deformable convolution.
For simplicity, CNet uses the  Pyramid, Cascading and Deformable convolutions (PCD) module proposed by EDVR~\cite{wang2019edvr} to align feature maps.
The detailed module structure is shown in Figure\ref{fig:vip}-(a).
The PCD module aligns the feature map of each frame $F_{t+i}^{L1}$ to the feature map of the middle frame $F_t^{L1}$.
In the alignment process, $F^{L1}_{[t-2:t+2]}$ is progressively convolved and down-sampled to obtain a small-scale feature maps $F^{L2}_{[t-2:t+2]}, F^{L3}_{[t-2:t+2]}$, and then the align is processed from $F^{L3}$ to $F^{L1}$ in a coarse-to-fine manner.
Align $F^{L1}_{[t-2:t+2]}$ with $F^{L1}_t$ respectively to obtain the aligned feature maps $F_{[t-2:t+2]}$.

\textbf{Inter-frame feature temporal fusion module (FTFM).}
The FTFM is used to fuse the feature maps from each frame to a compact and informative feature map for further process.
CVQENet directly uses the TSA module proposed by EDVR~\cite{wang2019edvr} for fusion, and the detailed structure is shown in Figure\ref{fig:vip}-(a).
The TSA module generates temporal attention maps through the correlation between frames and then performs temporal feature fusion through the convolutional layer.
Then, the TSA module uses spatial attention to further enhance the feature map.

\textbf{Decompression processing module (DPM).}
For the fused feature map, CVQENet uses the DPM module to remove artifacts caused by compression.
Inspired by RBQE~\cite{xing2020early}, DPM consists of a simple densely connected UNet, as shown in Figure\ref{fig:vip}-(b).
The $M_{i,j}$ cell contains Efficient Channel Attention (ECA)~\cite{wang2020eca} block, convolutional layer and residual blocks.
The ECA block performs adaptive feature amplitude adjustment through the channel attention mechanism.

\textbf{Feature quality enhancement module (FQEM).}
They add the output of DPM to $F^0_t$, then feed them into the feature quality enhancement module.
The shallow feature map $F^0_t$ contains a wealth of detailed information of the middle frame, which can help restore the middle frame.
The FQEM contains 20 stacked residual blocks without Batch Normalization (BN) layer to further enhance the feature maps and one convolutional layer (Conv) to generate the output frame~$\hat{I}_t$.

\subsubsection{Track 3}

Motivated by FastDVDnet~\cite{tassano2020fastdvdnet} and DRN~\cite{guo2020closed}, they propose the DUVE network for compressed video enhancement for Track 3, and the whole framework is shown in Figure~\ref{fig:vip2}-(a). It can be seen that given five consecutive compressed frames $I_{[t-2:t+2]}$, the goal of DUVE is to restore an uncompressed frame $\hat{O}_t$. Specifically, for five continuous input frames, each of the three consecutive images forms a group, so that the five images are overlapped into three groups $I_{[t-2:t]}$, $I_{[t-1:t+1]}$ and $I_{[t:t+1]}$. Then, the three groups are fed into Unet1 to get coarsely restored feature maps $\widetilde{F}_{t-1}^{c}$, $\widetilde{F}_{t}^{c}$ and $\widetilde{F}_{t+1}^{c}$, respectively. Considering the correlation between different frames and the current reconstruction frame $I_{t}$, the two groups of coarse feature maps $\widetilde{F}_{t-1}^{c}$ and $\widetilde{F}_{t+1}^{c}$ are filtered by nonlinear activation function $\sigma$ to get $F_{t-1}^{c}$ and $F_{t+1}^{c}$. Next, $F_{t-1}^{c}$, $\widetilde{F}_{t}^{c}$ and $F_{t+1}^{c}$ are concatenated along the channel dimension, and then pass through the channel reduction module to obtain fused coarse feature map $\hat{F}_{t}^{c}$. To further reduce compression artifacts, they apply UNet2 on $\hat{F}_{t}^{c}$ to acquire the fine feature map $\hat{F}_{t}^{f}$. Finally, a quality enhancement module takes the fine feature map to achieve the restored frame $\hat{O}_t$. The detailed architecture of Unet1 and Unet2 is shown in Figure~\ref{fig:vip2}-(b). In the proposed method, the mere difference between Unet1 and Unet2 is the number $n$ of Residual Channel Attention Blocks (RCAB)~\cite{zhang2018image}. The numbers n of RCABs in Unet1 and Unet2
are 6 and 10, respectively. The $L_{2}$ loss is utilized as the loss function.

\subsection{BLUEDOT Team}

\begin{figure}[!t]
\centering
\subfigure[The proposed framework.]{\includegraphics[width=\linewidth]{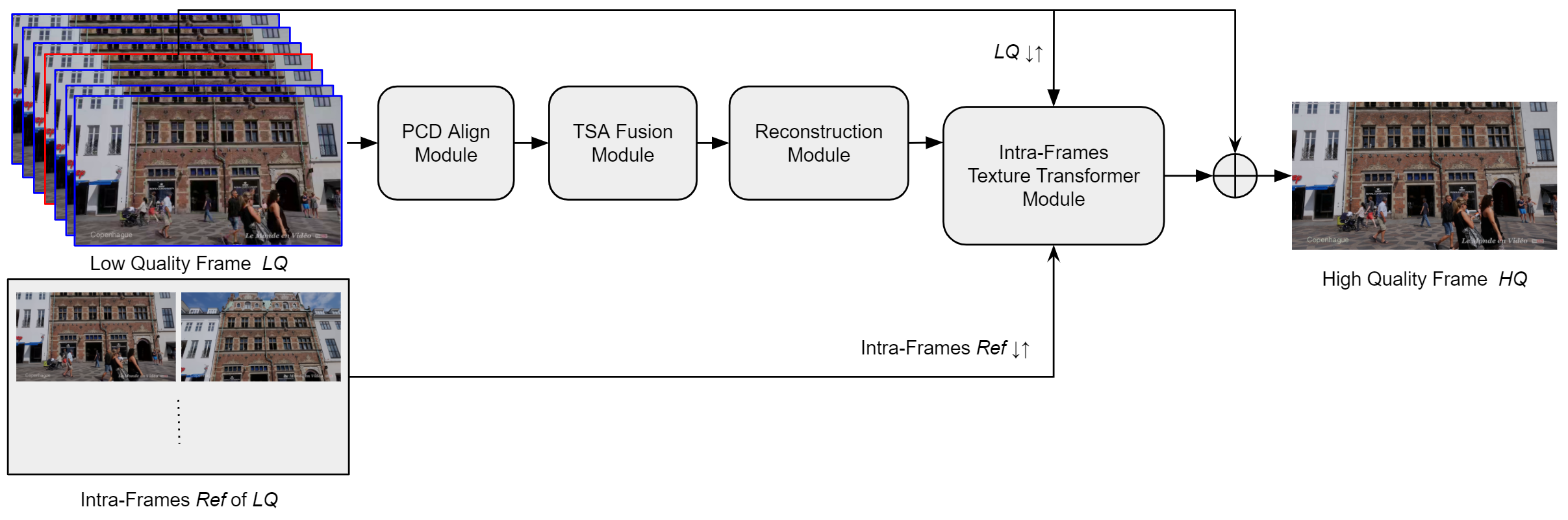}}
\subfigure[Intra-Frames Texture Transformer (IFTT).]{\includegraphics[width=.7\linewidth]{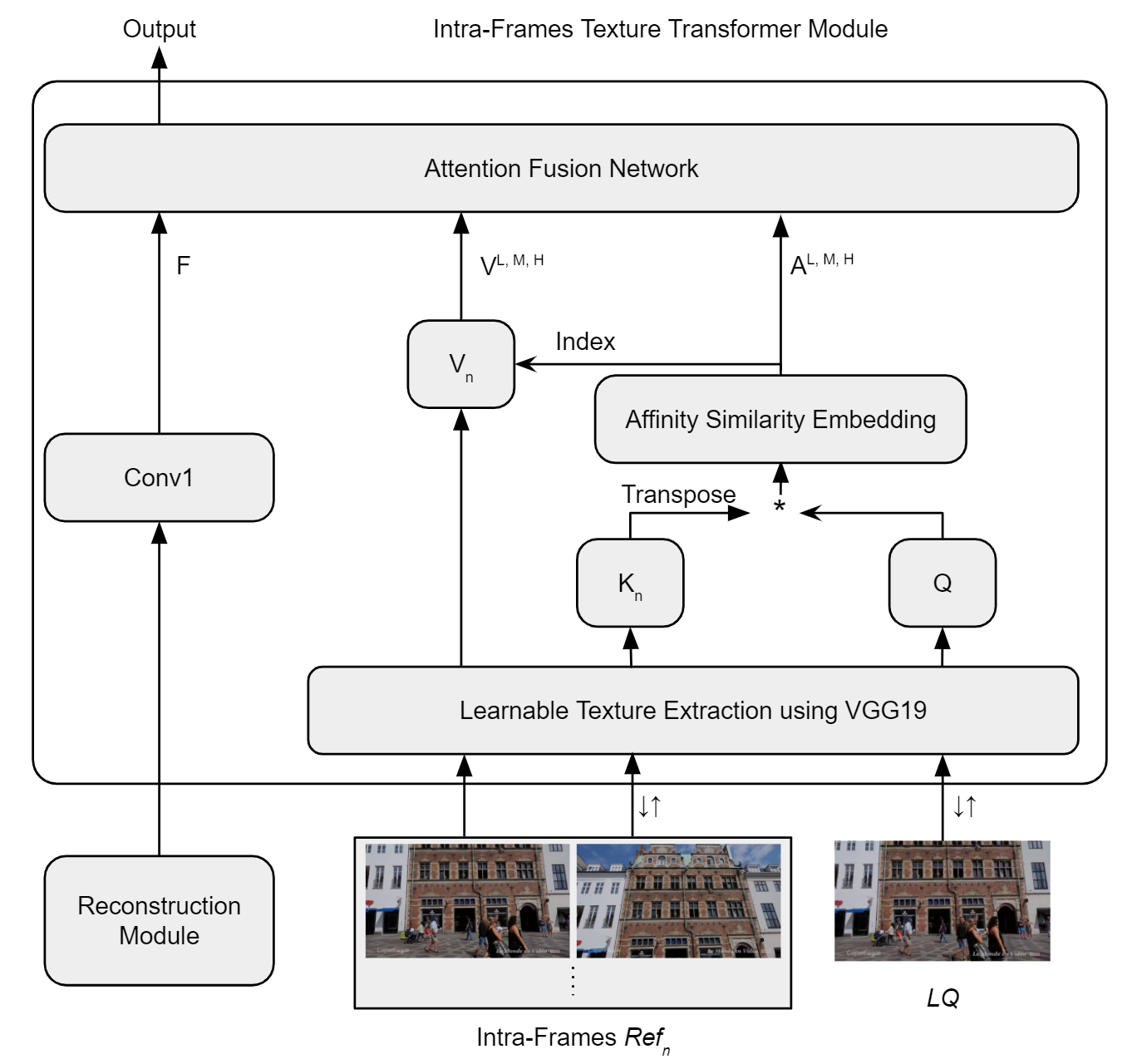}}
\subfigure[Attention fusion network in IFTT.]{\includegraphics[width=\linewidth]{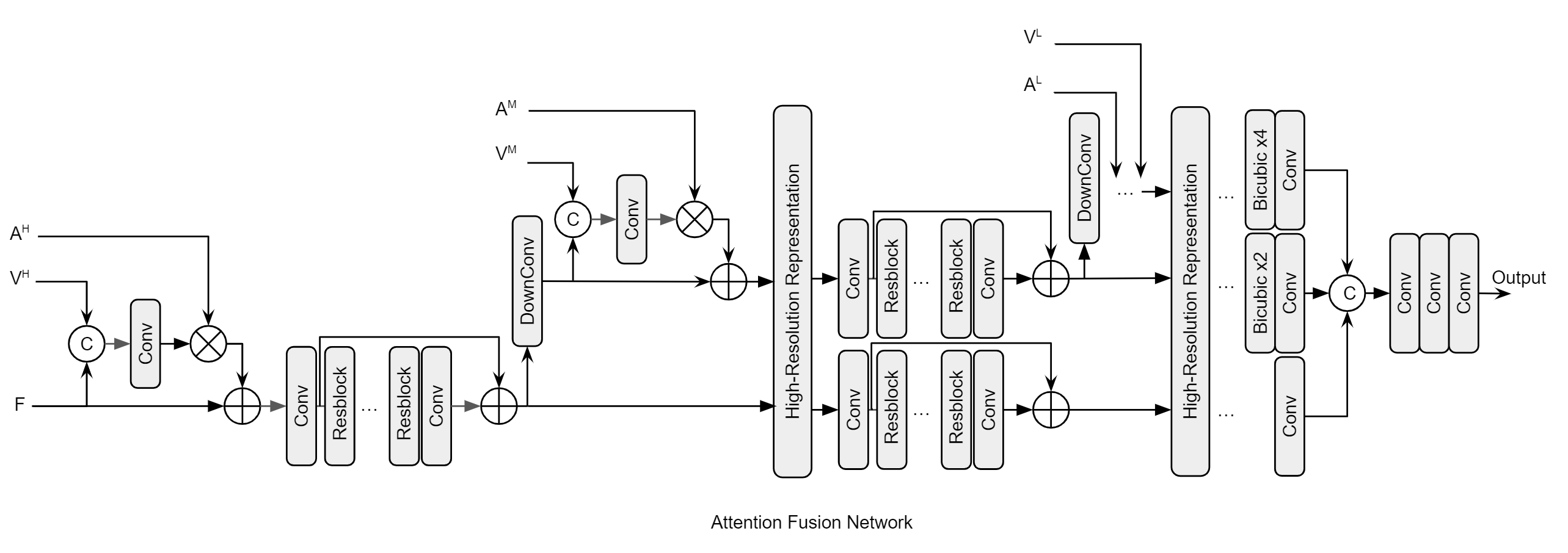}}
\caption{The proposed method of the BLUEDOT Team.}
\label{fig:blue}
\end{figure}

The proposed method is shown in Figure~\ref{fig:blue}, which is motivated by the idea that intra-frames usually have better video quality than inter-frames. It means that more information about the texture of the videos from intra-frames can be extracted. They build
and train a neural network built on EDVR~\cite{wang2019edvr} and TTSR~\cite{yang2020learning}. Relevances
of all intra-frames in the video are measured and one frame of the highest relevance with the current frame is embedded in the network. They carry out a two-stage training in the same network to obtain the video enhanced by the restored intra-frames. In the first stage, the model is learned by low-quality intra-frames. Then, in the second stage, the model is trained with the intra-frames enhanced by first-stage model.

\subsection{HNU\_CVers Team}

The HNU\_CVers Team proposes the patch-based heavy compression recovery model for Track 1.

\begin{figure}
    \centering
    \subfigure[Single-frame Residual Channel-Attention Network (RCAN\_A)]{\includegraphics[width=1\linewidth]{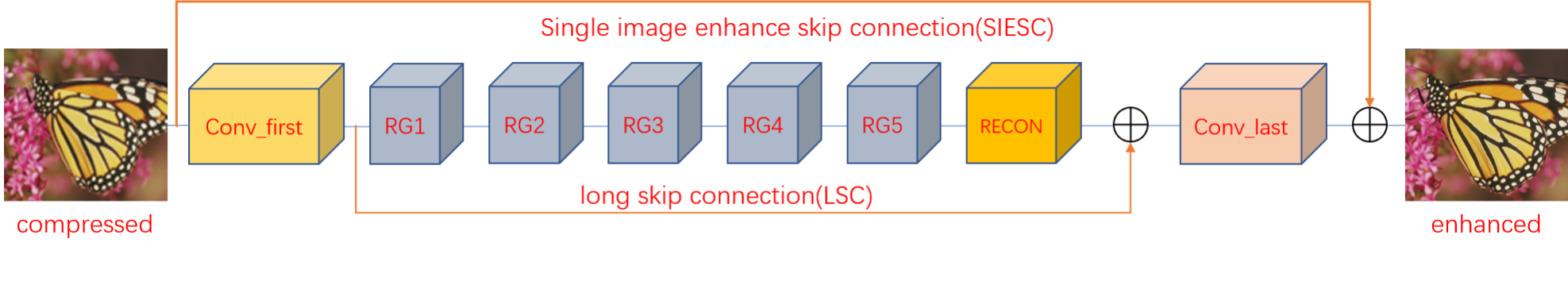}} 
    \subfigure[Multi-frame Residual Channel-Attention Video Network (RCVN)]{\includegraphics[width=1\linewidth]{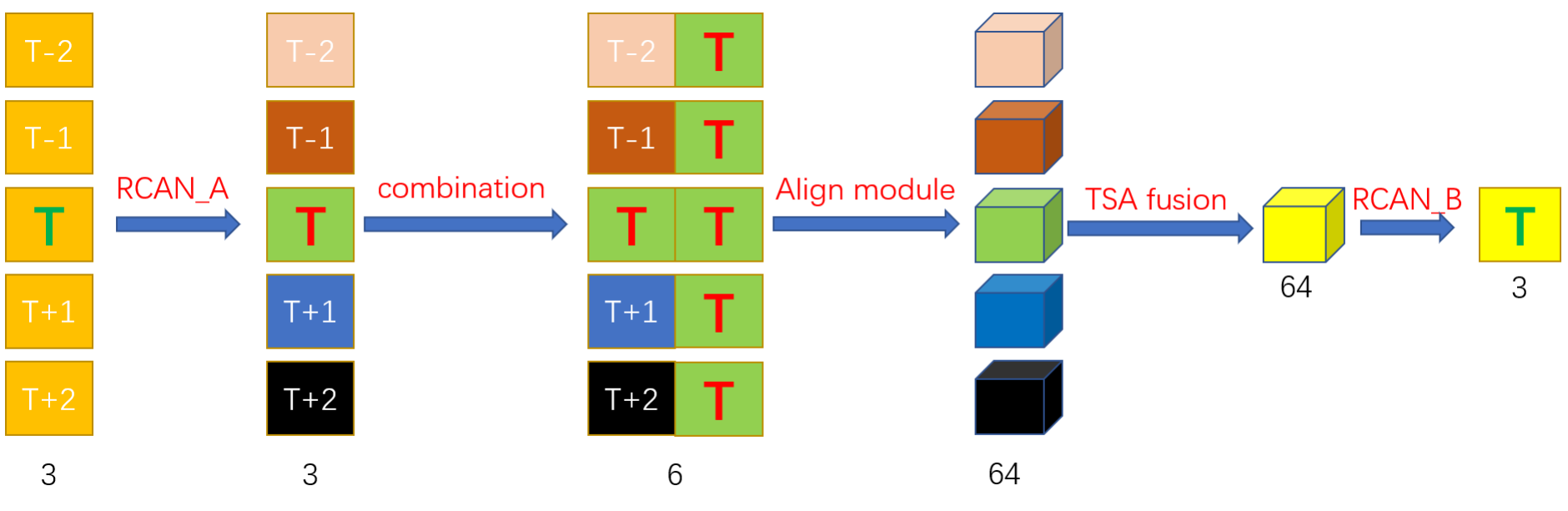}} 
    \caption{The proposed method of the HNU\_CVers Team.} 
    \label{HNU} 
\end{figure}

\textbf{Single-frame residual channel-attention network.} First, they remove the upsampling module from~\cite{zhang2018image} and add a skip connection to build a new model.
Different from~\cite{zhang2018image}, the RG (Residual Group) number is set as 5 and there are 16 residual blocks~\cite{lim2017enhanced} in each RG.
Each residual block is composed of $3\times3$ convolutional layers and ReLU with 64 feature maps. The single-frame model architecture is shown in Figure~\ref{HNU}-(a), called RCAN\_A. The model is trained with the $L1$ loss. 

\textbf{Multi-frame residual channel-attention video network.} They further make the model compact by taking as inputs five consecutive frames of images which are enhanced by
RCAN\_A. The multi-frame model architecture is shown in Figure~\ref{HNU}-(b). The five consecutive frames are 
stained with different colors in Figure~\ref{HNU}-(b) after enhanced by RCAN\_A. In order to mine the information of the 
consecutive frames, they combine the central frame with each frame. For
aligning, they design the [conv (64 features) + ReLU + Resbolcks (64 features) $\times$ 5] network as the
align network with the shared parameters for each combination. Immediately after it,
temporal and spatial attention are fused~\cite{wang2019edvr}. After getting a single frame feature map (colored yellow in Figure~\ref{HNU}-(b)), they adopt another model called RCAN\_B, which has the same structure as RCAN\_A. Finally, the restored RGB image is obtained through a convolution layer. The model is also trained with the $L1$ loss.

\textbf{Patch Integration Method (PIM).} They further propose a patch-based fusion model to strengthen the reconstruction ability of the multi-frame model. 
For a small patch, the model enhances the central part better than the edges.
Therefore, they propose feeding the overlapping patches to the proposed network. In the reconstructed patches, they remove the edges that overlap with the neighboring patches and only keep the high-confidence part in the center. 

\subsection{McEnhance Team}

\begin{figure}
\centering
\includegraphics[width=1\linewidth]{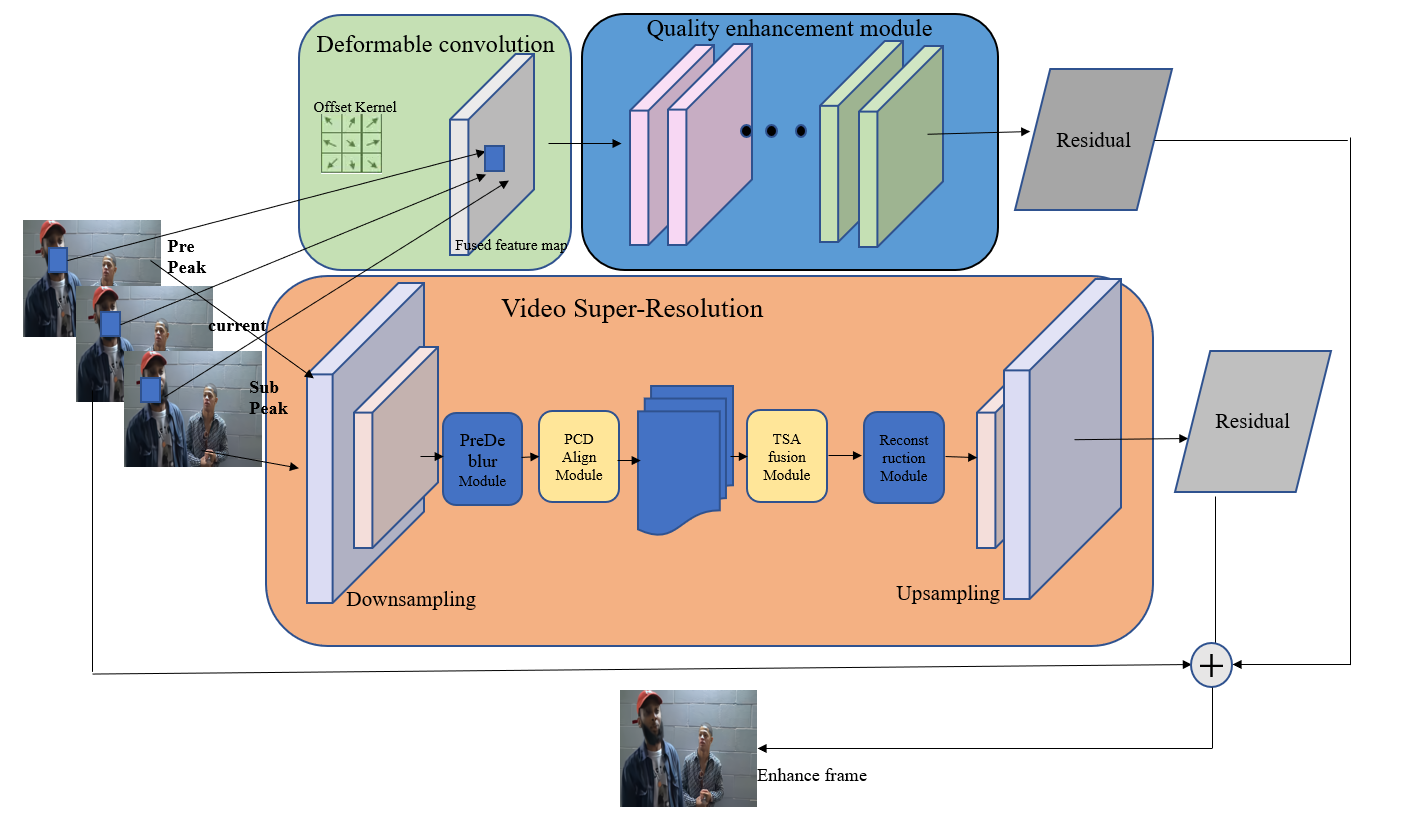}
\caption{The proposed method of the McEnhance Team.}
\label{fig:mc}
\end{figure}

The McEnhance Team combines video super-resolution technology~\cite{deng2020spatio,wang2019edvr} with multi-frame enhancement~\cite{yang2018multi, guan2019mfqe} to create a new end-to-end network, as illustrated in Figure~\ref{fig:mc}. First, they choose the current frame and its neighbor peak quality frames as the inputs. Then, they feed them to the deformable convolution network for alignment. As a result, complementary information from both target and reference frames can be fused by the operation. In the following, they feed them separately to the Quality Enhancement (QE) module~\cite{deng2020spatio} and the Temporal and Spatial Attention (TSA)~\cite{wang2019edvr} network. Finally, they add the two residuals to the compressed target frame to obtain the enhanced frame. There are two steps in the training stage. First, they calculate the PSNR of each frame in the training set and make labels of peak PSNR frames. Secondly, they feed the current frame and two neighbor peak PSNR frames to the proposed model.

\subsection{Ivp-tencent}

\begin{figure}[!t]
\centering
\includegraphics[width=1\linewidth]{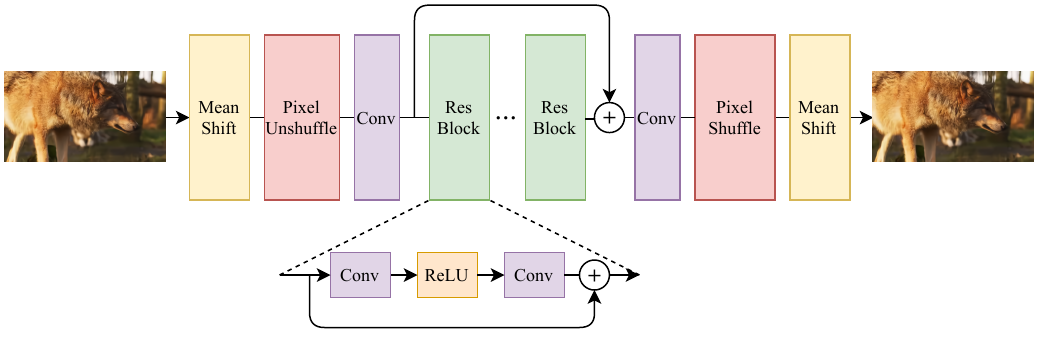}
\caption{The proposed BRNet of the Ivp-tencent Team.}
\label{fig:ivp}
\vspace{-1em}
\end{figure}

As Figure~\ref{fig:ivp} shows, the Ivp-tencent Team proposes a Block Removal Network
(BRNet) to reduce the block artifacts in compressed video for quality enhancement. Inspired by EDSR~\cite{lim2017enhanced} and FFDNet~\cite{zhang2018ffdnet}, the proposed BRNet first uses a mean shift module (Mean Shift) to normalize the input
frame, and then adopts a reversible down-sampling operation (Pixel Unshuffle) to
process the frame, which splits the compressed frame into four down-sampled sub-frames. Then, the sub-frames are fed into a convolutional network shown in Figure~\ref{fig:ivp}, in which they use 8 residual blocks. Finally, they use an up-sampling operation (Pixel Shuffle) and a mean shift module to reconstruct the enhanced frame. Note that, the up-sampling operation (Pixel Shuffle) is the inverse operation of the down-sampling operation (Pixel Unshuffle).
During the training phase, they crop the given compressed frames to $64\times64$ and feed them to the network with the batch size of 64. The Adam~\cite{kingma2014adam} algorithm is adopted to optimize the $L1$ loss, and learning rate is set to $10^{-4}$. The model is trained for 100,000 epochs.

The proposed BRNet achieves higher efficiency compared with EDSR and FFDNet. The reason is two-fold: First, the input frame is sub-sampled into
several sub-frames as inputs to the network. While maintaining the quality performance, the network parameters are effectively reduced and the receiving field of the network is increased. Second, by removing the batch normalization layer of the residual blocks, about 40\% of the memory usage can be saved during training.

\section*{Acknowledgments}

We thank the NTIRE 2021 sponsors: Huawei,
Facebook Reality Labs, Wright Brothers Institute, MediaTek, OPPO and ETH Zurich (Computer Vision Lab). We also thank the volunteers for the perceptual experiment of Track 2.

\section*{Appendix: Teams and affiliations}

\subsection*{NTIRE 2021 Challenge}

\noindent\textit{\textbf{Challenge:}}  

\noindent NTIRE 2021 Challenge on Quality Enhancement of Compressed Video

\noindent\textit{\textbf{Organizer(s):}} 

\noindent Ren Yang ({\texttt{ren.yang@vision.ee.ethz.ch}}), 

\noindent Radu Timofte ({\texttt{radu.timofte@vision.ee.ethz.ch}})

\noindent\textit{\textbf{Affiliation(s):}} 

\noindent Computer Vision Lab, ETH Zurich, Switzerland

\

\subsection*{Bilibili AI \& FDU Team}

\noindent\textit{\textbf{Title(s):}} 

\noindent Tracks 1 and 3: Spatiotemporal Model with Gated Fusion for Compressed Video Artifact Reduction

\noindent Track 2: Perceptual Spatiotemporal Model with Gated Fusion for Compressed Video
Artifact Reduction

\noindent\textit{\textbf{Member(s):}} 

\noindent Jing Liu (\texttt{liujing04@bilibili.com}), Yi Xu (\texttt{yxu17@fudan.edu.cn}), Xinjian Zhang, Minyi Zhao, Shuigeng Zhou

\noindent\textit{\textbf{Affiliation(s):}} 

\noindent Bilibili Inc.; Fudan University, Shanghai, China

\

\subsection*{NTU-SLab Team}

\noindent\textit{\textbf{Title(s):}} 

\noindent BasicVSR++: Improving Video Super-Resolution
with Enhanced Propagation and Alignment

\noindent\textit{\textbf{Member(s):}} 

\noindent Kelvin C.K. Chan (\texttt{chan0899@e.ntu.edu.sg}), Shangchen Zhou, Xiangyu Xu, Chen Change Loy

\noindent\textit{\textbf{Affiliation(s):}} 

\noindent  S-Lab, Nanyang Technological University, Singapore

\

\subsection*{VUE Team}

\noindent\textit{\textbf{Title(s):}} 

\noindent Tracks 1 and 3: Leveraging off-the-shelf BasicVSR for Video
Enhancement

\noindent Track 2: Adaptive Spatial-Temporal Fusion of Two-Stage Multi-Objective Networks

\noindent\textit{\textbf{Member(s):}} 

\noindent Xin Li (\texttt{lixin41@baidu.com}), He Zheng (\texttt{zhenghe01@baidu.com}), Fanglong Liu, Lielin Jiang, Qi Zhang, Dongliang He, Fu Li, Qingqing Dang

\noindent\textit{\textbf{Affiliation(s):}} 

\noindent Department of Computer Vision Technology (VIS), Baidu Inc., Beijing, China

\

\subsection*{NOAHTCV Team}

\noindent\textit{\textbf{Title(s):}} 

\noindent Multi-scale network with deformable temporal
fusion for compressed video restoration

\noindent\textit{\textbf{Member(s):}} 

\noindent Fenglong Song (\texttt{songfenglong@huawei.com}), Yibin Huang, Matteo Maggioni, Zhongqian Fu, Shuai Xiao, Cheng li, Thomas Tanay

\noindent\textit{\textbf{Affiliation(s):}} 

\noindent Huawei Noah's Ark Lab, Huawei Technologies Co., Ltd., Beijing, China

\

\subsection*{MT.MaxClear Team}

\noindent\textit{\textbf{Title(s):}} 

\noindent Enhanced EDVRNet for Quality Enhanced of Heavily Compressed Video

\noindent\textit{\textbf{Member(s):}} 

\noindent Wentao Chao (\texttt{cwt1@meitu.com}), Qiang Guo, Yan Liu, Jiang Li, Xiaochao Qu

\noindent\textit{\textbf{Affiliation(s):}} 

\noindent MTLab, Meitu Inc., Beijing, China

\

\subsection*{Shannon Team}

\noindent\textit{\textbf{Title(s):}} 

\noindent Disentangled Attention for Enhancement of Compressed Videos

\noindent\textit{\textbf{Member(s):}} 

\noindent Dewang Hou (\texttt{dewh@pku.edu.cn}), Jiayu Yang, Lyn Jiang, Di You, Zhenyu Zhang, Chong Mou

\noindent\textit{\textbf{Affiliation(s):}} 

\noindent Peking university, Shenzhen, China; Tencent, Shenzhen, China

\

\subsection*{Block2Rock Noah-Hisilicon Team}

\noindent\textit{\textbf{Title(s):}} 

\noindent Long Temporal Block Matching for Enhancement of Uncompressed Videos

\noindent\textit{\textbf{Member(s):}} 

\noindent Iaroslav Koshelev (\texttt{Iaroslav.Koshelev@skoltech.ru}), Pavel Ostyakov, Andrey Somov, Jia Hao, Xueyi Zou 

\noindent\textit{\textbf{Affiliation(s):}} 

\noindent Skolkovo Institute of Science and Technology, Moscow, Russia; Huawei Noah's Ark Lab; HiSilicon (Shanghai) Technologies CO., LIMITED, Shanghai, China

\

\subsection*{Gogoing Team}

\noindent\textit{\textbf{Title(s):}} 

\noindent Two-stage Video Enhancement Network for Different QP Frames

\noindent\textit{\textbf{Member(s):}} 

\noindent Shijie Zhao (\texttt{zhaoshijie.0526@bytedance.com}), Xiaopeng Sun, Yiting Liao, Yuanzhi Zhang, Qing Wang, Gen Zhan, Mengxi Guo, Junlin Li

\noindent\textit{\textbf{Affiliation(s):}} 

\noindent ByteDance Ltd., Shenzhen, China

\

\subsection*{NJU-Vision Team}

\noindent\textit{\textbf{Title(s):}} 

\noindent Video Enhancement with Progressive Alignment and Data Augmentation

\noindent\textit{\textbf{Member(s):}} 

\noindent Ming Lu (\texttt{luming@smail.nju.edu.cn}), Zhan Ma

\noindent\textit{\textbf{Affiliation(s):}} 

\noindent School of Electronic Science and Engineering, Nanjing University, China

\

\subsection*{BOE-IOT-AIBD Team}

\noindent\textit{\textbf{Title(s):}} 

\noindent Fully 3D-Convolutional MultiGrid-BackProjection Network

\noindent\textit{\textbf{Member(s):}} 

\noindent Pablo Navarrete Michelini (\texttt{pnavarre@boe.com.cn})

\noindent\textit{\textbf{Affiliation(s):}} 

\noindent BOE Technology Group Co., Ltd., Beijing, China

\

\subsection*{VIP\&DJI Team}

\noindent\textit{\textbf{Title(s):}}

\noindent Track 1: CVQENet: Deformable Convolution-based Compressed Video Quality Enhancement Network

\noindent Track 3: DUVE: Compressed Videos Enhancement with Double U-Net

\noindent\textit{\textbf{Member(s):}} 

\noindent Hai Wang (\texttt{wanghai19@mails.tsinghua.edu.cn}), Yiyun Chen, Jingyu Guo, Liliang Zhang, Wenming Yang

\noindent\textit{\textbf{Affiliation(s):}} 

\noindent Tsinghua University, Shenzhen, China; SZ Da-Jiang Innovations Science and Technology Co., Ltd., Shenzhen, China

\

\subsection*{BLUEDOT Team}

\noindent\textit{\textbf{Title(s):}} 

\noindent Intra-Frame texture transformer Network for compressed video enhancement

\noindent\textit{\textbf{Member(s):}} 

\noindent Sijung Kim (\texttt{jun.kim@blue-dot.io}), Syehoon Oh

\noindent\textit{\textbf{Affiliation(s):}} 

\noindent BLUEDOT, Seoul, Republic of Korea

\

\subsection*{HNU\_CVers Team}

\noindent\textit{\textbf{Title(s):}} 

\noindent Patch-Based Multi-Frame Residual Channel-Attention Networks For Video Enhancement

\noindent\textit{\textbf{Member(s):}} 

\noindent Yucong Wang (\texttt{1401121556@qq.com}), Minjie Cai

\noindent\textit{\textbf{Affiliation(s):}} 

\noindent College of Computer Science and Electronic Engineering, Hunan University, China

\

\subsection*{McEnhance Team}

\noindent\textit{\textbf{Title(s):}} 

\noindent Parallel Enhancement Net

\noindent\textit{\textbf{Member(s):}}

\noindent Wei Hao (\texttt{haow6@mcmaster.ca}), Kangdi Shi, Liangyan Li, Jun Chen

\noindent\textit{\textbf{Affiliation(s):}} 

\noindent McMaster University, Ontario, Canada

\

\subsection*{Ivp-tencent Team}

\noindent\textit{\textbf{Title(s):}} 

\noindent BRNet: Block Removal Network 

\noindent\textit{\textbf{Member(s):}} 

\noindent Wei Gao (\texttt{gaowei262@pku.edu.cn}), Wang Liu, Xiaoyu Zhang, Linjie Zhou, Sixin Lin, Ru Wang

\noindent\textit{\textbf{Affiliation(s):}} 

\noindent School of Electronic and Computer Engineering, Shenzhen Graduate School, Peking University, China; Peng Cheng Laboratory, Shenzhen, China; Tencent, Shenzhen, China

{\small
\bibliographystyle{ieee_fullname}
\bibliography{egbib}
}

\end{document}